\documentclass[aps,pre,amsmath,twocolumn,superscriptaddress,showpacs]{revtex4-2}
\usepackage{calc}
\usepackage{amsmath}
\usepackage{amsfonts}
\usepackage{amssymb}
\usepackage{graphicx}
\usepackage{hyperref}
\usepackage{soul}
\usepackage{mathtools}
\usepackage{verbatim}
\usepackage{epstopdf}
\usepackage{subfigure}
\usepackage{latexsym}
\usepackage{dcolumn}
\usepackage{epsf}
\usepackage{float}
\usepackage[table]{xcolor}
\usepackage{multirow}
\usepackage[toc]{appendix}
\usepackage{color} 
\usepackage{graphicx}% Include figure files
\usepackage{graphicx,epstopdf}
\usepackage{xcolor}
\usepackage{dcolumn}% Align table columns on decimal point
\usepackage{bm}% bold math
\usepackage{mathrsfs} % script-like, curvy letters.
\usepackage{mathrsfs}
\begin{document}
\title{Cluster formation due to repulsive spanning trees in attractively coupled networks}

\author{Sayantan Nag Chowdhury}
\email{jcjeetchowdhury1@gmail.com}
\affiliation{Department of Environmental Science and Policy, University of California, Davis, California 95616, USA}
\affiliation{Physics and Applied Mathematics Unit, Indian Statistical Institute, 203 B. T. Road, Kolkata 700108, India}
%\author{Francesco Sorrentino}
%\affiliation{University of New Mexico, Albuquerque New Mexico 87106, USA}
\author{Md Sayeed Anwar}
\affiliation{Physics and Applied Mathematics Unit, Indian Statistical Institute, 203 B. T. Road, Kolkata 700108, India}
\author{Dibakar Ghosh}
%\email{dibakar@isical.ac.in}
\affiliation{Physics and Applied Mathematics Unit, Indian Statistical Institute, 203 B. T. Road, Kolkata 700108, India}

%\subject{epidemic spreading, complex network}

%\keywords{Complex network, epidemic spreading, final outbreak size, test-kit}
%\thanks{These two authors contributed equally}
	
%\thanks{Corresponding Auhtor: Chittaranjan Hens}
%\email{chittaranjanhens@gmail.com}

%%%% Abstract text to be placed here %%%%%%%%%%%%
\date{\today}
\begin{abstract}
	
	Ensembles of coupled nonlinear oscillators are a popular paradigm and an ideal benchmark for analyzing complex collective behaviors. The onset of cluster synchronization is found to be at the core of various technological and biological processes. The current literature has investigated cluster synchronization by focusing mostly on the case of attractive coupling among the oscillators. However, the case of two coexisting competing interactions is of practical interest due to their relevance in diverse natural settings, including neuronal networks consisting of excitatory and inhibitory neurons, the coevolving social model with voters of opposite opinions, ecological plant communities with both facilitation and competition, to name a few. In the present article, we investigate the impact of repulsive spanning trees on cluster formation within a connected network of attractively coupled limit cycle oscillators. We successfully predict which nodes belong to each cluster and the emergent frustration of the connected networks independent of the particular local dynamics at the network nodes. We also determine local asymptotic stability of the cluster states using an approach based on the formulation of a master stability function. We additionally validate the emergence of solitary states and antisynchronization for some specific choices of spanning trees and networks.
	
\end{abstract}

\maketitle

%\begin{quotation}
%	{\bf  Many research efforts have been devoted to studying synchronization among interactive oscillatory elements in attractively coupled networks. More research is needed to understand synchronization in connected networks with both attractive and repulsive interactions. We consider a connected undirected network of coupled identical oscillators. We propose a universal method based on the identification of a spanning tree within the connected network by which we are able to divide the population of coupled oscillators into two clusters whenever we pass a negative coupling of adequate strength through that spanning tree. It is possible to determine the members of each cluster without numerical simulations by partitioning the vertex of the repulsive spanning tree. In such networks with mixed attractive and repulsive interactions, the notion of link frustration is used to determine the appropriate repulsive coupling strength to avoid multistability. Our findings are further validated by showing the stability of the antisynchronous cluster solutions. Our work provides a deeper understanding of networks of with both attractive and repulsive interactions.} %positively and negatively coupled oscillators involving asymmetrical choices and clustering in real-world settings, such as the brain's neuronal network and the social processes of opinion dynamics.}
%
%\end{quotation}

\section{Introduction} \label{introduction}

\par Due to practical applications in biological and physical systems, the emergence of collective order \cite{strogatz2004sync,arenas2008synchronization,ghosh2022synchronized,anwar2021relay,lei2005synchronization,shajan2021enhanced,chatterjee2022controlling,hramov2004approach,sorrentino2007network,frolov2021extreme,lei2006global,mirzaei2023synchronization,sorrentino2012synchronization,anwar2022intralayer} in systems consisting of many interacting dynamical elements is a thoroughly studied topic. Apart from complete synchronization \cite{threshold_2019,syncpik,chowdhury2019convergence}, different kinds of collective states, viz.\ relay synchronization \cite{rakshit2021relay,leyva2018relay,banerjee2012enhancing}, chimeras \cite{majhi2019chimera,parastesh2021chimeras,khaleghi2019chimera}, solitary states \cite{maistrenko2014solitary,majhi2019solitary,teichmann2019solitary},  antisynchronization \cite{zhang2004chaotic,kim2003anti,hu2005adaptive}, cluster synchronization \cite{lu2010cluster,belykh2008cluster,pecora2014cluster}, extreme events \cite{NAGCHOWDHURY20221,mishra2020routes,chowdhury2021extreme1}, splay states \cite{strogatz1993splay,zou2009splay,doi}, amplitude and oscillation death states \cite{dixit2021emergent,ad_report,dixit2021dynamic}, have gained significant attention among the scientific community. All the studies have shown that these emergent states depend crucially not only on the intrinsic properties of the individual oscillators but also on the nature of the interaction among them. Large part of the literature considers a generic setting consisting of a network, where each node of the network is an oscillator and each link represents an interaction between pairs of oscillators. Both the network topology as well as the coupling play a vital role in the origin of a specific collective dynamical state in ensembles of coupled oscillators.

\par  In what follows, we often refer to coupling among the network nodes as either attractive or repulsive. Attractive coupling is conducive to stabilizing the oscillators on the same time evolution while repulsive coupling is not. The simplistic approach of considering only attractive coupling may not be sufficient to capture the essence of several real-world systems. There is only a limited literature on signed networks consisting of attractive and repulsive couplings \cite{majhi2020perspective}. However, understanding such systems is essential for their relevant applications from neuroscience with inhibitory and excitatory connections \cite{soriano2008development,vogels2009gating} to opinion formation \cite{do2013phase,mikaberidze2024consensus} in social networks.

\par In the present article, instead of choosing only positive (attractive) coupling strength that generally facilitates the minimization of the phase difference among the coupled oscillators, we consider a mixed coupling of positive and negative interactions within the same network. The presence of co-existing positive and negative couplings opens a new avenue for studying various natural settings. For example, Refs.\ \cite{chowdhury2020distance,chowdhury2019extreme} recently showed that the coupled oscillators synchronize despite the presence of both positive-negative couplings, thanks to the temporal connectivity of the mobile agents. The interplay between such competing interactions can produce a novel $\pi$-state as described in Refs.\ \cite{hong2011conformists,yuan2018periodic,Sar_2022,hong2011kuramoto}. Ref.\ \cite{leyva2006sparse} reported that such repulsive coupling can easily enhance the synchronization in a small world network among attractively coupled nonidentical dynamical units. The interested reader may refer to the excellent mini-review \cite{majhi2020perspective} on such systems consisting of both attractive and repulsive interactions. 

\par
\begin{figure}[htp]
	\centerline{\includegraphics[width=0.25\textwidth]{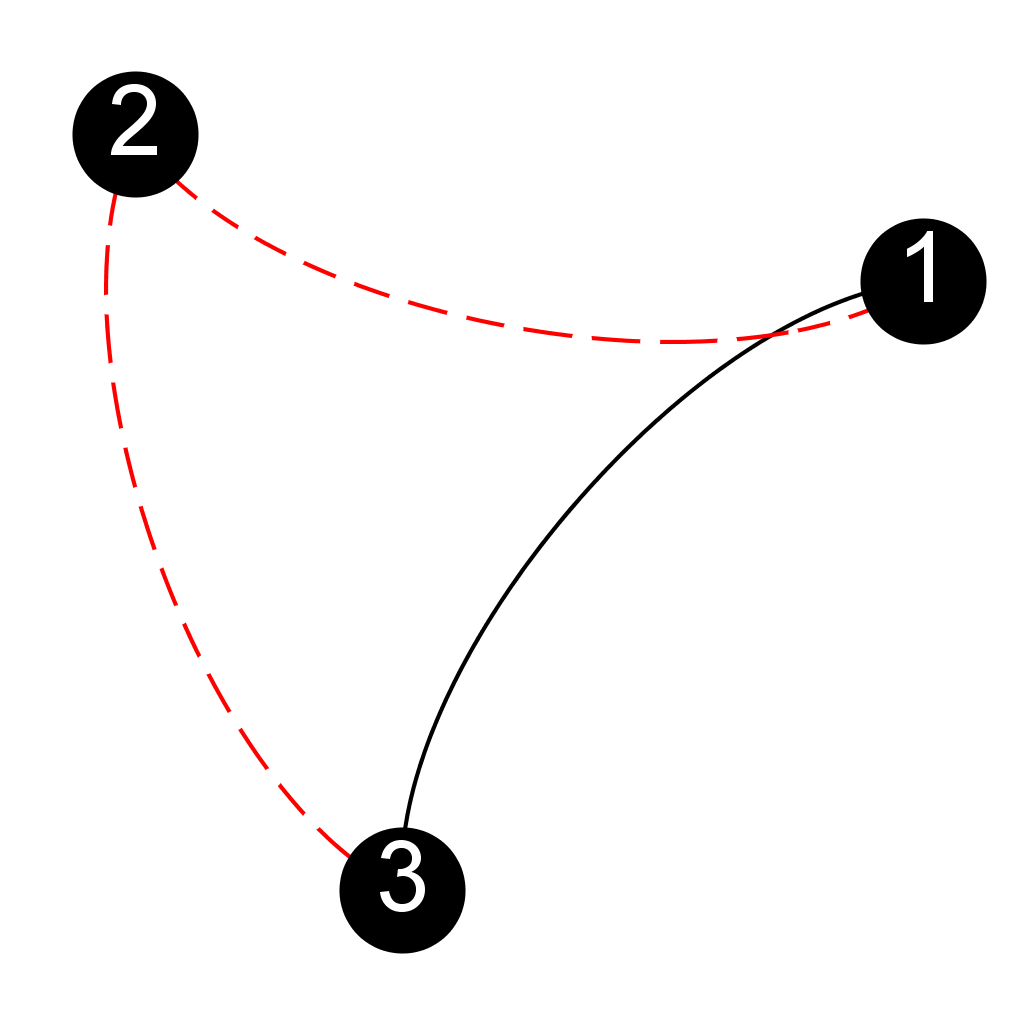}}
	\caption{{\bf The complete graph $K_3$}: We consider here the most straightforward non-bipartite connected network with $N=3$ nodes. The red dashed arcs represent one of the available spanning trees of $K_3$. This spanning tree is isomorphic to the other two available spanning trees.}
	\label{Picture1}
\end{figure}

 The article \cite{chowdhury2021antiphase} showed that whenever repulsive coupling of adequate strength is passed through a dedicated subgraph in a bipartite network, the system may settle down on a state of antiphase synchronization. This corresponds to a dynamical state in which all of the adjacent oscillators maintain a phase difference of $\pi$, i.e., the whole system splits into two distinct groups or clusters. The oscillators within the same cluster possess almost identical phases, while the phase difference between the two clusters is approximately $\pi$. 
Some research has indicated that antiphase synchronization may be commonly observed in cortical neural networks \cite{li2011organization}. However, one can not attain antiphase synchronization in any network with arbitrary structure. For example, let us consider a ring network of $N=3$ oscillators as depicted in Fig.\ \eqref{Picture1}. %Let $\theta_i$ be the intrinsic phase of the $i$-th oscillator. Thus, 
If the system exhibits antiphase synchronization, the phase difference between connected neighbors is $\pi$. If the instantaneous phase of oscillator-$1$ is $\alpha$ at a particular time, then the instantaneous phases of oscillators-$2$ and $3$ will be $\alpha \pm \pi$ at that time. But oscillators-$2$ and $3$ are connected too. So their phase difference should be $\pm \pi$. Then, we can not achieve antiphase synchronization in such network. In fact, the presence of an odd cycle does not allow the system to settle down on a state of antiphase synchronization. Reference \cite{chowdhury2021antiphase} showed that one needs a network without odd cycles to entertain antiphase synchronization. {\color{black}Thus, the focus in Ref.\ \cite{chowdhury2021antiphase} was on networks without odd cycles, which were found to be necessary for antiphase synchronization, particularly in bipartite networks.
Our manuscript extends this investigation to encompass both bipartite and non-bipartite networks, where odd cycles may indeed be present. This extension broadens the scope of our study and introduces novel insights into the dynamics of synchronization in connected networks with diverse topological structures.} %Hence, one needs a network without any odd cycles to entertain antiphase synchronization. This understanding is also consistent with the findings of the Ref.\ \cite{chowdhury2021antiphase}.

\par The present article identifies a suitable subgraph within a larger network, such that whenever we pass repulsive coupling of appropriate strength between the edges of that subgraph, the onset of two clusters with $\pi$ phase difference is observed. Such repulsive subgraph can be found in any connected graph. Moreover, one can quickly identify the members of each cluster using the vertex decomposition of this repulsive path independent of the local dynamics. Previous work on competitive positive-negative coupling \cite{majhi2020perspective} has focused on coupled phase oscillators. Phase oscillators with sine coupling, like the Kuramoto and Kuramoto-Sakaguchi models, have received much attention thanks to the possibility of the analytical tractability of such systems. %Previous work on competitive positive-negative coupling \cite{majhi2020perspective} has focused on coupled phase oscillators. Phase oscillators with the sine coupling like the Kuramoto and Kuramoto-Sakaguchi model have received much attention, thanks to the possibility of analyzing such systems.
However, the case of oscillators characterized by an amplitude and a phase is more complicated. We place identical Stuart-Landau oscillators \cite{kuramoto2003chemical} on top of each node. We show the emergence of a particular state in which the oscillators within the same cluster possess almost identical phases and amplitudes; nevertheless, the phases of oscillators from different clusters differ by an angle $\pi$.  

\par In Sec.\ \eqref{The model}, we describe our designed model with mixed positive-negative coupling. We further present necessary information about the subgraphs and the dynamical states used in the rest of the article. Sec.\ \eqref{Results} presents our key findings with a few undirected networks using numerical simulations. We also present comparisons between the emergent frustration \cite{chowdhury2020effect} in the network and along the chain using different repulsive subgraphs of the same length. We can successfully predict these frustrations theoretically without detailed information on the local dynamics. We further study the stability of the synchronized clusters in networks of attractive-repulsively coupled limit cycle oscillators. To ensure our discoveries are accurate, we test them on different kinds of small networks. Some of these networks are complete networks consisting of three or four nodes, while others have random structures having six, eight, or ten nodes (Please see the Appendix \ref{10_node}-\ref{6_node_2} for the results on connected networks with random structures). These smaller networks confirm that our findings are accurate and make it easier for us to explain them. However, we also test our proposals on a much larger complete network with $100$ nodes. Section \eqref{Conclusion} summarizes our findings.

\section{The model} \label{The model}
 We consider the following equation describing the dynamics of a graph $\mathcal{G}$ of $d$-dimensional coupled limit cycle oscillators,
\begin{equation} \label{eq:1}
\begin{array}{lcl}
	\dot{{\bf x}}_i=f({\bf x}_i)+ \sum_{j=1}^{N} D_{ij} H({\bf x}_i,{\bf x}_j),i=1,2,...,N.
\end{array}
\end{equation}
Here, $f({\bf x}_i): \mathbb{R}^d \rightarrow \mathbb{R}^d$ governs the time evolution of the oscillator-$i$ with state vector ${\bf x}_i (t)$. The interaction function $H({\bf x}_i, {\bf x}_j)$ is a vector coupling from $\mathbb{R}^d$ to $\mathbb{R}^d$. {\color{black}We select both $f$ and $H$ as odd functions. The rationale behind these selections is elaborated upon explicitly in the subsequent subsection \eqref{K4}.} The matrix $D={[D_{ij}]}_{N \times N}$ represents the connectivity structure of the  graph $\mathcal{G}$ where the entries of the matrix $D$ are equal to $D_{ij}=\epsilon_A B_{ij}+\epsilon_R C_{ij}$, where $\epsilon_A>0$ is the attractive coupling strength, and $\epsilon_R<0$ is the repulsive coupling strength. We consider two different types of node-to-node interactions: attractive and repulsive. The attractive information is passed through a subgraph $\mathcal{G}_{1}$ of $\mathcal{G}$, characterized by the adjacency matrix $B={[B_{ij}]}_{N \times N}$. %$\epsilon_A>0$ is the attractive coupling strength.
Besides, we choose another particular subgraph $\mathcal{G}_{2}$ associated with the adjacency matrix $C={[C_{ij}]}_{N \times N}$ through which we advance the repulsive coupling strength $\epsilon_R<0$. Thus, the sum of these two binary matrices $B={[B_{ij}]}_{N \times N}$ and $C={[C_{ij}]}_{N \times N}$ provide a new matrix $A={[A_{ij}]}_{N \times N}$ which represents the adjacency matrix of the whole network $\mathcal{G}$. %with unit coupling strengths (i.e., $\epsilon_{A}=\epsilon_{R}=1$).
Further, we assume the whole network to be connected and the interaction between any two oscillator is bi-directional and simple, i.e, there is either attractive or repulsive link between them.

\par For illustration, in Fig.\ \eqref{Picture1} we present the complete graph $K_3$ composed of three-node connected to each other. The red dashed arcs in that figure reflect the repulsive edges of the chosen regular graph $K_3$. The black arc is used to represent the attractive edge through which we pass the positive coupling strength. Thus, for this case 

\begin{equation}
	A = \begin{bmatrix}
		0 & 1 & 1 \\
		1 & 0 & 1 \\
		1 & 1 & 0
	\end{bmatrix}
	= \begin{bmatrix}
		0 & 0 & 1 \\
		0 & 0 & 0 \\
		1 & 0 & 0
	\end{bmatrix}
	+
	\begin{bmatrix}
		0 & 1 & 0 \\
		1 & 0 & 1 \\
		0 & 1 & 0
	\end{bmatrix}
\end{equation}

%$A$=\[
%\begin{bmatrix}
%0      & 1 & 1 \\
%1      & 0 & 1 \\
%1 & 1 & 0
%\end{bmatrix}
%=
%\begin{bmatrix}
%0      & 0 & 1 \\
%0      & 0 & 0\\
%1       & 0 & 0
%\end{bmatrix}
%+
%\begin{bmatrix}
%0      & 1 & 0 \\
%1      & 0 & 1\\
%0       & 1 & 0
%\end{bmatrix}
%\]
The first matrix on the right-hand side is $B$, and the second one represents the matrix $C$. The zero (one) entries indicate the absence (presence) of a link in either graph.  Throughout the article, we use the terms \{networks, graphs\}, \{edges, links, arcs\}, and \{vertices, nodes\} interchangeably.
\begin{figure*}[htp]
\centerline{\includegraphics[width=1.0\textwidth]{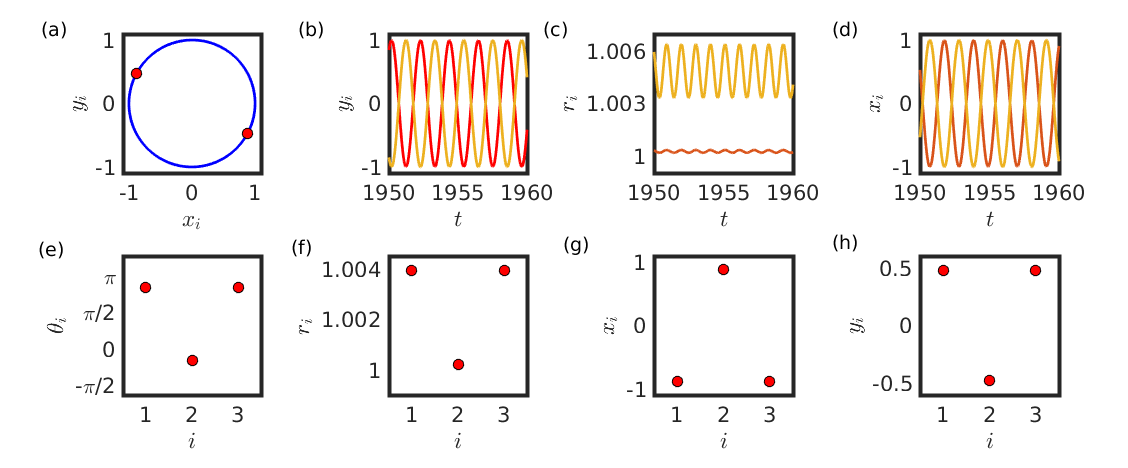}}
\caption{{\bf Emergence of cluster synchronization in $K_3$}: The introduction of the repulsive path (red dashed links) in $K_3$ as shown in Fig.\ \eqref{Picture1} induces cluster synchrony. The system splits into two distinct groups. The vertex decomposition $V_1=\{1,3\}$ and $V_2=\{2\}$ of the repulsive spanning tree remains consistent with the cluster synchronization, where oscillators-$1$ and $3$ belong to a group and the remaining oscillator-$2$ lies on the other cluster. The phase difference between these two clusters is $\pi$. Subfigure (a) shows the phase portrait, and subfigures (b-d) reveal the temporal evolution of $y_i$, $r_i$, and $x_i$, respectively. (e-h) The second row displays the snapshots at a specific iteration after the transients. For further information, please see the main text.}
\label{Picture2}
\end{figure*}
\par Now, the presence of both competing couplings creates a tug of war in the whole network. We will measure this by adapting the concept of frustration \cite{chowdhury2020effect} given by the following expression
\begin{equation}\label{F}
F=\Big\langle \frac{1}{L}\sum_{i<j}^{}A_{ij}\left(1+\cos(\theta_i-\theta_j)\right) \Big\rangle_t,
\end{equation}
where 
\begin{equation}
L=\dfrac{1}{2} \sum_{i=1}^{N} \sum_{j=1}^{N} A_{ij}=\dfrac{1}{2} \sum_{i=1}^{N} \sum_{j=1}^{N} \big( B_{ij}+C_{ij} \big)
\end{equation}
is the total number of links of the graph $\mathcal{G}$. $\theta_i$ is the intrinsic phase of the $i$-th oscillator. Here, ${\langle \cdot \cdot \cdot \rangle}_t$ indicates the time average taken over sufficiently long iterations after the initial transient. Moreover, we are also interested in calculating the frustration 
\begin{equation}
F_{chain}=\Big\langle \dfrac{1}{N-1}\sum_{i=1}^{N-1}\left(1+\cos(\theta_i-\theta_{i+1})\right) \Big\rangle_t,
\end{equation}
considering the localized frustration between the oscillator-$i$ and oscillator-$(i+1)$ for $i=1,2,3,\cdots,(N-1)$. When the measure $F$ is zero, the phase difference among two connected neighbors is the highest as $\theta_i-\theta_j = \pm \pi$. This corresponds to the presence of antiphase synchronization between two adjacent oscillators, while $F=2$ indicates the presence of inphase synchronization for which $\theta_i-\theta_j = 0$ \cite{chowdhury2020effect}. Note that the phase difference between neighbors will contribute in the measure $F$ only when there exists a link between those two oscillators (i.e., \ $A_{ij}=1$). However, the calculation of $F_{chain}$ only takes into account the phase difference between any two oscillators-$i$ and $i+1$ (not necessarily, they are neighbors), viz.\ $i$ and $(i+1)$ for $i=1,2,3,\cdots,(N-1)$. It will be zero if the oscillators ${1,2,3,\cdots, N}$ divide in two clusters, and these two clusters are opposite in phase with each other ($|\theta_i-\theta_{i+1}| = \pi$ for $i=1,2,3,\cdots,(N-1)$). Neither one of these two measures incorporates information regarding the amplitude of the oscillators.

\par The primary goal of this article is to trace out a path that exists in any connected graph so that we pass repulsive coupling strength through the selected subgraph and the ensemble of oscillators split into exactly two synchronized groups of oscillators, irrespective of the chosen underlying network. Further, we want to predict the members of both clusters without investigating the local dynamics. In what follows, we choose the paradigmatic Stuart-Landau (SL) limit cycle oscillator as the state dynamics of individual limit cycle oscillator given by 
\begin{equation} \label{eqL}
f({\bf x}_i)=\left(
\begin{array}{c}
	\left[1-\left({x_i}^2+{y_i}^2\right)\right]x_i-\omega_i y_i\\\\
	\left[1-\left({x_i}^2+{y_i}^2\right)\right]y_i+\omega_i x_i\\
\end{array}
\right). \\
\end{equation}

We set the intrinsic frequency $\omega_i=\omega=3.0$ to be identical for all the oscillators and choose the coupling function $H({\bf x}_i,{\bf x}_j)=[{x}_j + {x}_i, 0]^T$ which resembles a rescaled pairwise mean-field interaction through the first state variable of the SL oscillator. For the numerical simulation, we derive the instantaneous phase of each oscillator through the principal value (i.e., $\tan\theta_i= \frac{y_i}{x_i}$) of the argument of the complex number $z_i=x_i+\sqrt{-1}y_i=r_i exp(\sqrt{-1}\theta_i)$. $r_i=\sqrt{x_i^2+y_i^2}$ is the distance from the pole (the origin) in the polar coordinate plane. All of our numerical simulations are performed using the Runge–Kutta–Fehlberg method with fixed integration time-step $\delta t=0.01$.

\begin{figure*}[htp]
	\centerline{\includegraphics[width=0.65\textwidth]{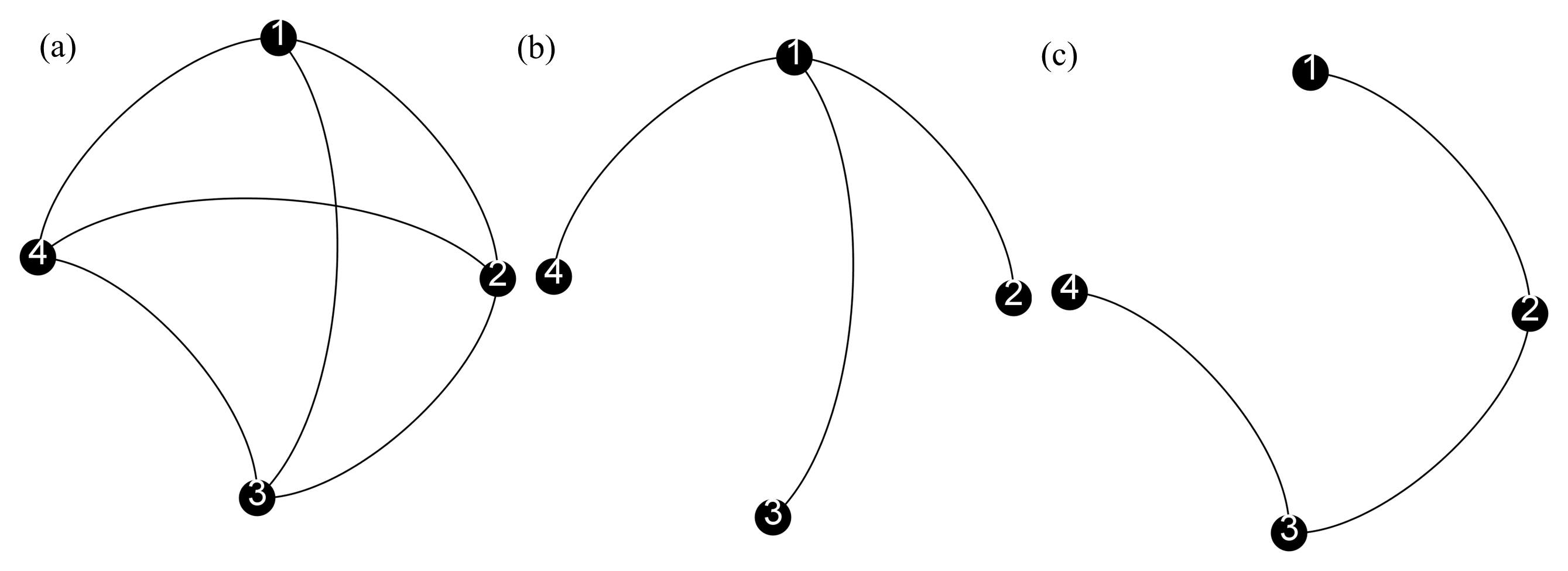}}
	\caption{{\bf $K_4$ and its non-isomorphic spanning trees}: The complete graph $K_4$ with $4$ vertices possesses $16$ spanning trees, out of which the number of non-isomorphic spanning trees is $2$. We draw $K_4$ in subfigure (a), and the other two subfigures in (b-c) represent those two non-isomorphic spanning trees. This figure is drawn using the software Gephi \cite{bastian2009gephi}.}
	\label{Picture3}
\end{figure*}

\section{Results} \label{Results}

\par Before discussing our key findings, we must discuss three critical things to understand the presented phenomenon in a better way. Firstly, we restrict our study mainly to non-bipartite networks. Our results are also valid for any bipartite networks, as long as the networks remain connected. The results for bipartite networks have already been discussed in Refs.\  \cite{chowdhury2020effect,chowdhury2021antiphase}. A bipartite graph is such that the set of vertices can be partitioned into two disjoint sets such as there is no link between the members of the same set \cite{sorrentino2007network}. References \cite{chowdhury2020effect,chowdhury2021antiphase} have shown that antiphase synchronization can be achieved in biparitite graphs based on the choice of the initial conditions and of the repulsive coupling strength. However, non-bipartite graphs contain odd cycles and hence, cannot achieve such an arrangement. The fundamental goal of the present study is to identify a subgraph within any connected network so that whenever we pass sufficiently strong negative coupling strength through that path, the coupled oscillators converge to two $\pi$-distant apart clusters. In this article, in order to achieve such novel states, we set $|\epsilon_A| << |\epsilon_R|$. Other choices of $\epsilon_A$ and $\epsilon_R$ may not be able to overcome the emergent multistability in the system \eqref{eq:1}. Particularly, multistability becomes more common with increasing network size. Such multistable signature is common among repulsive oscillators as described in Ref.\ \cite{chowdhury2020effect,levnajic2011emergent}. We also confirm it using the link frustration $F$ given in Eq.\ \eqref{F}. We identify huge fluctuations in every figure of the variation of $F$  depending on the choice of $\epsilon_R<0$ and initial conditions. In what follows we fix $\epsilon_A$ such that when all the links are attractive the network achieves in-phase synchronization.  We pass the negative coupling through any of the spanning trees (explicitly discussed in the next subsections) to achieve two approximately $\pi$-distant apart cluster states. On the other hand, we set $|\epsilon_A| << |\epsilon_R|$ in the most cases. We explicitly discuss all these components of our results so the readers can quickly identify the reason behind the choice of weaker attractive coupling strength in magnitude than the repulsive coupling strength. %Still, throughout the study, we fix $\epsilon_A =\epsilon_R>0$ initially in such strength so that we can achieve almost close to in-phase synchronization.
{\color{black} Throughout the study, we maintain $\epsilon_A = \epsilon_R > 0$ as our initial configuration to achieve nearly in-phase synchronization. Note that $F=2$ reveals in-phase synchronization, and we plot the variation of $F$ in various figures to confirm this in-phase synchronization for $\epsilon_A =\epsilon_R>0$. Subsequently, we vary $\epsilon_R$ from positive to negative values while holding $\epsilon_A > 0$ fixed and constant. After achieving the desired collective state, we determine the range of $\epsilon_R$ within which the collective state is observable and locally stable. Subsequently, we set the coupling strength to a fixed value, $\epsilon_R < 0$, from that determined range.}  
%The last component of this discussion is to mention that even if we choose $\epsilon_A=\epsilon_R>0$, our system \eqref{eq:1} may not be able to maintain complete synchronization depending on network topology and coupling function. To prove this, let us assume that the system \eqref{eq:1} entertains complete synchronization beyond a suitable choice of coupling strength $\epsilon_A=\epsilon_R>0$. Since, $\epsilon_A=\epsilon_R>0$, then Eq.\ \eqref{eq:1} transforms to
%
%\begin{equation} 
%	\begin{array}{lcl}
%		\dot{{\bf x}}_i=f({\bf x}_i)+ \epsilon_A \sum_{j=1}^{N} A_{ij} H({\bf x}_i,{\bf x}_j),i=1,2,...,N.
%	\end{array}
%\end{equation}
%
%Since the oscillators synchronize as per our assumption, then we can choose any two arbitrary oscillators, namely $k$ and $l$-th oscillators from the population, and we find ${\bf x}_l={\bf x}_k$ after an initial transient. These two oscillators then maintain the following differential equations respectively
%
%\begin{equation} 
%	\begin{array}{lcl}
%		\dot{{\bf x}}_l=f({\bf x}_l)+ \epsilon_A \sum_{j=1}^{N} A_{lj} H({\bf x}_l,{\bf x}_j), \\
%		\dot{{\bf x}}_k=f({\bf x}_k)+ \epsilon_A \sum_{j=1}^{N} A_{kj} H({\bf x}_k,{\bf x}_j). 
%	\end{array}
%\end{equation}
%
%Since ${\bf x}_l={\bf x}_k$, then substracting these two equations, we obtain
%
%\begin{equation} 
%	\begin{array}{lcl}
%		 \sum_{j=1}^{N} (A_{lj}-A_{kj}) H({\bf x}_l,{\bf x}_j)=0. 
%	\end{array}
%\end{equation}

\begin{figure*}[htp]
	\centerline{\includegraphics[width=0.65\textwidth]{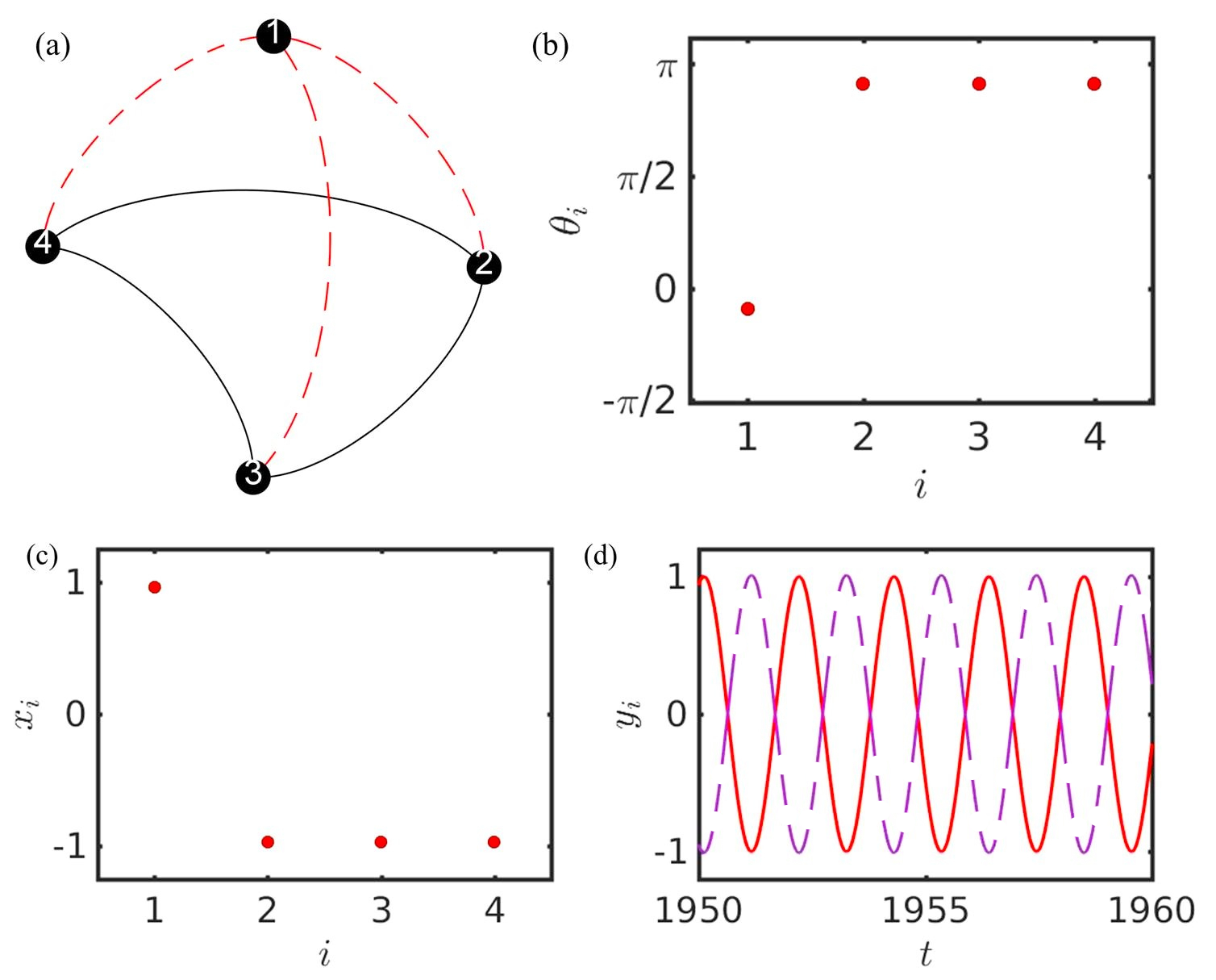}}
	\caption{{\bf Predicting cluster synchronization in undirected and connected network $K_4$}: (a) We pass the repulsive coupling through the spanning tree (red dashed arcs) shown in Fig.\ \eqref{Picture3} (b). The vertex decomposition of this spanning tree allows predicting which oscillators belong to which clusters. The black arcs depict the links through which we pass the positive coupling strength ${\epsilon}_{A}=0.01$. Subfigure (b) assures the phase difference between these two clusters is exactly $\pi$. Subfigure (c) further confirms oscillator-$1$ belongs to one cluster, and the other three oscillators lie on another group. The temporal evolution of $y_i$ in subfigure (d) reveals the trajectories are oscillating in the limit cycle regime maintaining cluster synchronization with $\pi$ phase difference.}
	\label{Picture4}
\end{figure*}

\begin{figure*}[htp]
\centerline{\includegraphics[width=0.55\textwidth]{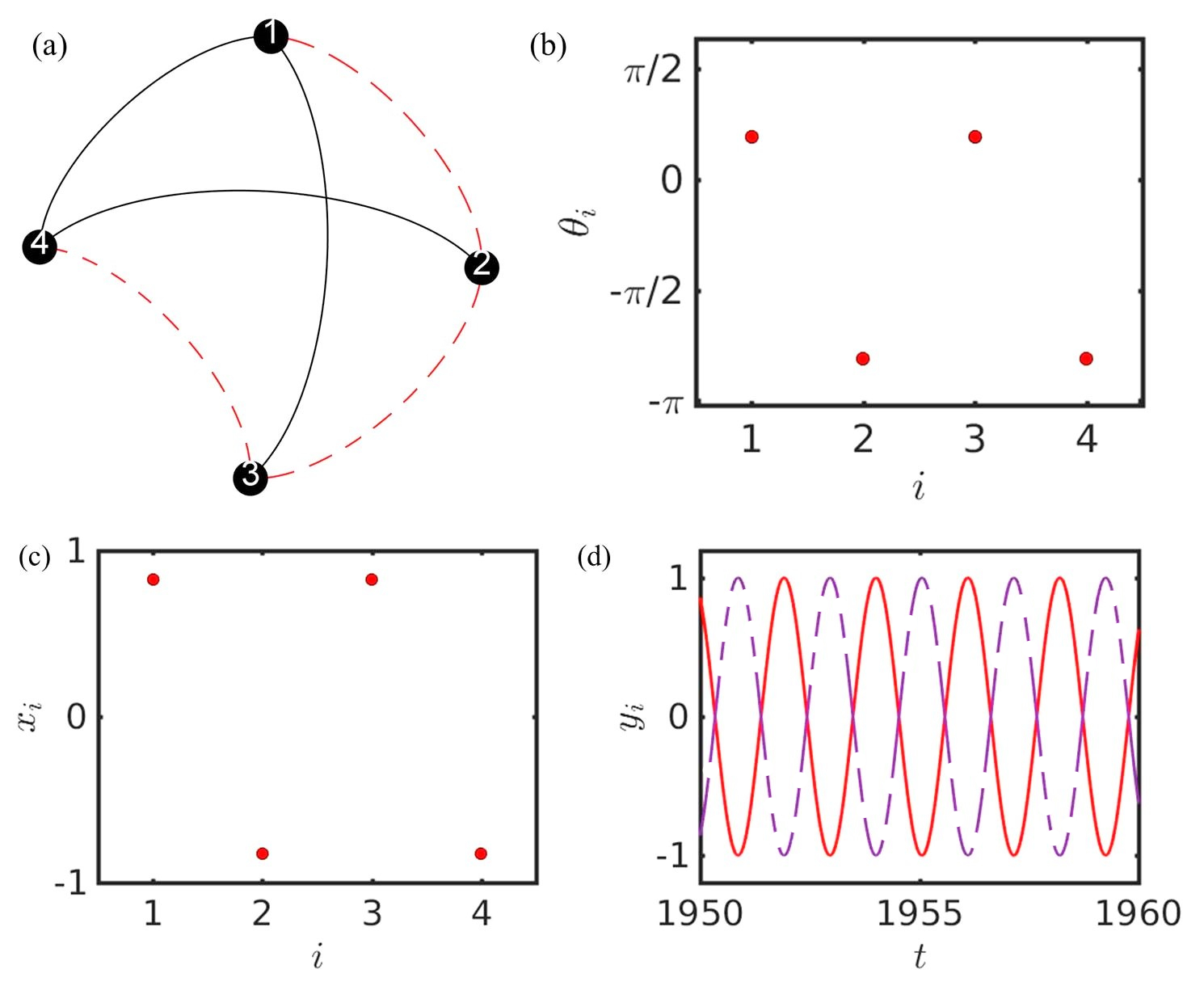}}
\caption{{\bf Identifying the members of the clusters in $K_4$}: (a) We choose a spanning tree (red dashed arcs) as shown in Fig.\ \eqref{Picture3} (c) of $K_4$. We pass the repulsive coupling with $\epsilon_R=-0.1$ through this spanning tree. (b) Oscillators-$1$ and $3$ lies in one cluster, and the other two oscillators belong to the other cluster. Moreover, both of these clusters maintain a phase difference of $\pi$. (c) The vertex partitioning of the repulsive spanning tree predicts the members of the clusters without investigating the local dynamics. The snapshot of $x_i$ at a particular time after the transient validates our claim here. (d) The oscillating temporal evolution of the trajectories contemplates a phase difference of $\pi$ between these two clusters.}
\label{Picture5}
\end{figure*}

%Note that we arbitrarily choose two oscillators with indices $k$ and $l$. Thus, we arrive at the necessary conditions for achieving complete synchronization in any connected networks, namely $\sum_{j=1}^{N} A_{lj}=\sum_{j=1}^{N} A_{kj}$; i.e., the degree of each node must be equal. That implies we need a regular network for obtaining complete synchronization unless $ H({\bf x}_l,{\bf x}_j)=0$. $ H({\bf x}_l,{\bf x}_j)=0$ implies the coupling
%function H vanishes after achieving complete synchronization. Nevertheless, we choose the coupling function $H({\bf x}_i,{\bf x}_j)=[{x}_j + {x}_i, 0]^T$ that does not vanish after the oscillators of the network evolve synchronously. Thus, we must need a regular network to manifest complete synchronization. Therefore, even if we choose $\epsilon_A =\epsilon_R>0$ sufficiently large, the system \eqref{eq:1} may not achieve complete synchronization unless the chosen connected network is regular.

\subsection{The complete graph $\mathbf{K_3}$}
\subsubsection{Spanning trees}
\par To begin with, we choose the network $\mathcal{G}$ to be complete graph $K_3$. This is the smallest non-bipartite graph with $3$ vertices. %We start with a non-bipartite graph intentionally because bipartite graphs are capable of generating antiphase synchronization under suitable conditions \cite{chowdhury2021antiphase}. Such $2$-colorable graphs \cite{asratian1998bipartite,deo2017graph} (depending on intrinsic properties of the individual oscillators, the network topology, the nature of the coupling, etc.) can easily induce two almost in-phase synchronized clusters, and those two clusters maintain a phase difference of $\pi$. Hence, we purposefully choose a non-bipartite graph with only $3$-nodes. Also, 
In the case of a connected network with only two vertices,
the only connection can either be attractive or repulsive, hence we discard the case of $N=2$.

% we can introduce only a single type of coupling strength, either the attractive or the repulsive. Hence, we discard the results for two coupled systems. 

\begin{figure*}[htp]
\centerline{\includegraphics[width=0.65\textwidth]{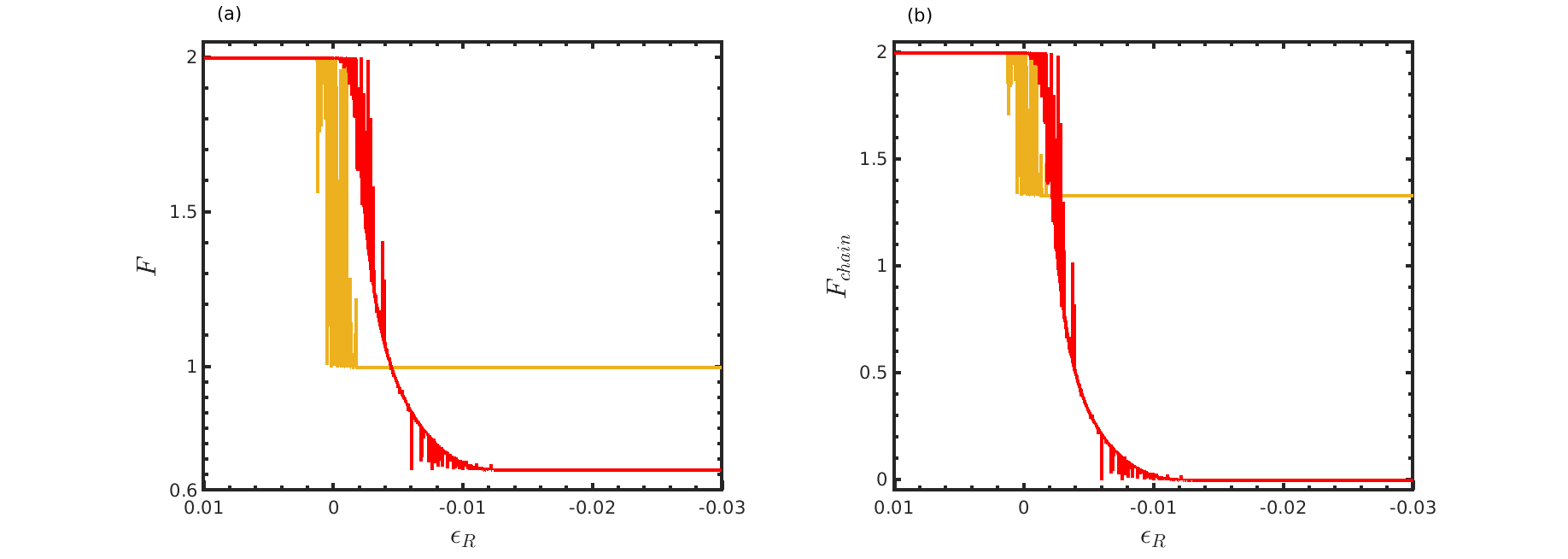}}
\caption{{\bf Numerically calculated frustration through two different repulsive spanning trees in $K_4$}: We change the value of repulsive coupling strength $\epsilon_R$ in the range $[-0.03,0.01]$, starting from $10^{-2}$ with fixed step-length $=-10^{-5}$. The initial conditions are drawn for each $\epsilon_R$ randomly within $[-1,1] \times [-1,1]$. The red (dark gray) line represents the result when the repulsive path is chosen as drawn in Fig.\ \eqref{Picture3}(c). The yellow (light gray) line portrays the impact of the spanning tree as shown in Fig.\ \eqref{Picture3}(b). Here, the attractive coupling strength is kept fixed at $\epsilon_A=0.01$. $F$ deals with the frustration of the whole network, while the measure $F_{chain}$ contains the information of the frustration only through the chain. The spanning tree in \eqref{Picture3}(c) is more effective than the spanning tree in \eqref{Picture3}(b) in terms of attaining the minimum value of $F$ and $F_{chain}$. {\color{black}Every data point in this figure represents a single realization of the process.}}
\label{Picture6}
\end{figure*}

\par For passing the repulsive interaction, we choose a subgraph of $K_3$ in such a way that this subgraph contains all the vertices of $K_3$. The existence of such a subgraph in any network is guaranteed using the connectedness of the underlying network. If the chosen connected network does not contain any cycle, it will serve as its own such subgraph. The presence of a cycle within the connected graph provides us the opportunity to delete one link from each present cycle and construct a new connected subgraph containing all the nodes of the underlying connected network. Hence, our study is restricted within the domain of the connected network. Generally, in the existing literature, such a subgraph is known as a spanning tree \cite{graham1985history,deo2017graph}, and its edges are referred to as branches \cite{deo2017graph}. In our investigation, we choose all the repulsive interactions through the branches of a spanning tree, and the remaining links are subjected to attractive interactions. More precisely, the subgraph through which we pass the repulsive interactions is a spanning tree of the whole network.

\subsubsection{Emergence of Clusters}
\par Typically, a connected graph may contain more than one spanning tree. For instance, this complete graph $K_3$ possesses exactly three spanning trees as per Cayley's formula \cite{aigner1998proofs}. But, all these three spanning trees are isomorphic to one another. Hence, we choose a specific spanning tree consisting of two links $1-2$ and $2-3$. We pass the repulsive coupling strength $\epsilon_R=-0.1$ through this spanning tree (see the red dashed arcs in Fig.\ \eqref{Picture1})and the positive coupling strength $\epsilon_A=0.01$ through the remaining link (the black arc of $K_3$). The equation \eqref{eq:1} yields
\begin{equation}\label{solitary}
\begin{split}
	\dot{{\bf x}}_1=f({\bf x}_1)+ \epsilon_R H({\bf x}_1,{\bf x}_2)+ \epsilon_A H({\bf x}_1,{\bf x}_3), \hspace{0.2cm}\\
	\dot{{\bf x}}_2=f({\bf x}_2)+ \epsilon_R [H({\bf x}_2,{\bf x}_1)+ H({\bf x}_2,{\bf x}_3)], \hspace{0.2cm}\\
	\dot{{\bf x}}_3=f({\bf x}_3)+ \epsilon_R H({\bf x}_3,{\bf x}_2)+ \epsilon_A H({\bf x}_3,{\bf x}_1). \hspace{0.2cm}
\end{split}
\end{equation}
After the initial transient, we find ${\bf{x}}_1={\bf{x}}_3 \approx - {\bf{x}}_2$ (see Fig. \ref{Picture2}). Due to our choice of repulsive spanning tree, oscillator-$2$ experiences only the effect of negative coupling strength through the undirected edges $1-2$ and $2-3$. Thus, our chosen coupling function $H$ provides $H({\bf x}_2,{\bf x}_1)$ and $H({\bf x}_2,{\bf x}_3)$ approximately equal to $[0, 0]^T$ after the initial transient. Therefore, the equation \eqref{solitary} can be rewritten as, 
\begin{equation}\label{solitary1}
\begin{split}
	\dot{{\bf x}}_1 \approx f({\bf x}_1)+ \epsilon_A H({\bf x}_1,{\bf x}_3), \hspace{0.2cm}\\
	\dot{{\bf x}}_2 \approx f({\bf x}_2), \hspace{0.2cm}\\
	\dot{{\bf x}}_3 \approx f({\bf x}_3)+ \epsilon_A H({\bf x}_3,{\bf x}_1), \hspace{0.2cm}
\end{split}
\end{equation}
from which we see that a synchronized solution is possible in which ${\bf{x}}_1(t)={\bf{x}}_3(t)$.

The existence of positive coupling strength between the oscillators-$1$ and $3$ allows them to maintain a coherent oscillation. Moreover, we choose the repulsive coupling strength from the saturated domain for fixed attractive coupling strength $\epsilon_A=0.01$, such that the system settles down to these two clusters for all initial conditions chosen from $[-1,1] \times [-1,1]$. The temporal evolutions of the state variables in Figs. \ref{Picture2}(b)-\ref{Picture2}(d) and the snapshot of phase distribution in Fig. \ref{Picture2}(e) attest both these clusters remain approximately $\pi$-distance apart. The numerical derivation gives $F \approx 0.666$, and $F_{chain}=0$. %{\textit{Can the members of these clusters be predicted before the numerical simulation? Can we predict the $F$ and $F_{chain}$ prior to the numerical calculation?} 

\begin{figure*}[htp] 
	\centerline{
		\includegraphics[scale=0.2]{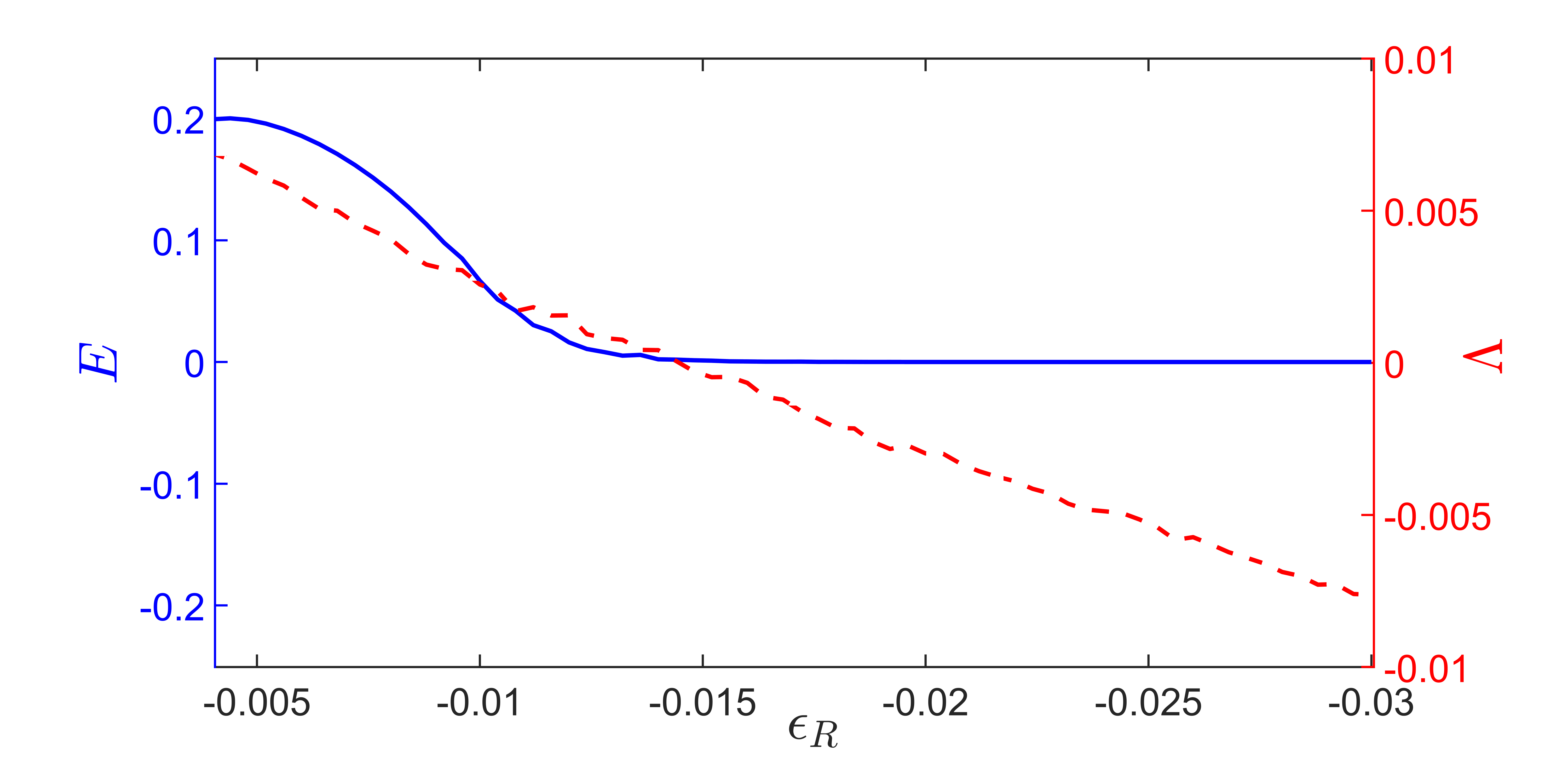}}
	\caption{{\bf Stability of cluster synchronization state:} Cluster synchronization error $E$ (blue line) and the maximum Lyapunov exponent transverse to the cluster synchronization manifold $\Lambda$ (red dashed line) as a function of repulsive coupling strength $\epsilon_{R}$. The corresponding network structure is illustrated in Fig.\ \eqref{Picture5}(a). The attractive coupling strength $\epsilon_{A}$ is kept fixed at $0.01$. For further information, please see the main text.}
	\label{sw_random}
\end{figure*}

\subsubsection{Cluster Member Prediction}
\par We observe that the repulsive spanning tree is a bipartite graph, and thus the set of vertices of this repulsive subgraph can be decomposed into two disjoint and independent sets $V_1=\{1,3\}$ and $V_2=\{2\}$. Noticeably, the set of vertices of the repulsive spanning tree can be uniquely partitioned into two distinct sets, and we notice the emergence of two different clusters in Fig.\ \eqref{Picture2}. This vertex partitioning is consistent with the members of the clusters too. This inspection offers an efficient, effortless method of determining the members of both groups in advance of the numerical calculation. For the theoretical prediction purpose, without loss of any generality, we assign the phase value of $\alpha$ (say) to each element of the set $V_1$ and allocate the phase value of $\alpha + \pi$ to each oscillator of $V_2$. The edges between these two sets $V_1$ and $V_2$ will not contribute anything to the measure $F$, as the phase difference between these sets is $\pi$ as per our allocation. The set $V_1$ contains two distinct elements, and a link exists between these two oscillators. Since, as per our assignment, oscillators belonging to the same set are in-phase synchronized, thus the term $A_{13}\left(1+\cos(\theta_1-\theta_3)\right)$ will contribute $2$. Hence, our predicted $F$ will be $\dfrac{2}{3}$, which matches excellently with our numerical analysis. The denominator of $F$ represents the number of links $L=3$ in the whole network $K_3$. Similarly, we can predict the value $F_{chain}$. We find the theoretically predicted value of $F_{chain}$ is zero for the chosen network $K_3$. We also verify our prediction of $F$ and $F_{chain}$ in $K_2$. We detect $F=0$ and $F_{chain}=0$ in $K_2$ as per our theoretical as well as numerical assessment. This $F=0$ confirms the occurrence of antiphase synchronization in the two repulsively-coupled system (figures are not shown here). However, the presence of an odd cycle in $K_3$ restricts $F$ to diminish to zero. Still, the arrangement of repulsive coupling leads to the emergence of a two-cluster state, where the oscillators of different clusters exhibit a phase-difference of $\pi$.

\subsection{The complete graph $\mathbf{K_4}$} \label{K4}
\subsubsection{Impact of Various Spanning Trees}
\par To validate the effectiveness of our described method for predicting the members of the clusters, we choose a different globally connected network $K_4$ with $N=4$ vertices and $L=6$ edges (see Fig.\ \eqref{Picture3} (a)). Cayley's formula {\color{black}predicts} this network possesses exactly $16$ spanning trees. We consider only two spanning trees of $K_4$, which is shown in Fig.\ \eqref{Picture3} (b-c). These two spanning trees are non-isomorphic to each other as two isomorphic graphs must contain the same number of vertices of the same degree. Moreover, all other $14$ spanning trees are isomorphic to any one of these considered spanning trees {\color{black}(Please refer to the Appendix \ref{k4_spanning} for further explanations)}. 

\subsubsection{Solitary state}

\par We first choose the spanning tree shown in Fig.\ \eqref{Picture3} (b) and pass the repulsive coupling strength $\epsilon_R=-0.1$ through this subgraph. The vertex partitioning of the set of vertices of the chosen spanning tree gives the disjoint sets $V_1=\{1\}$ and $V_2=\{2,3,4\}$. Our arrangement of attractive-repulsive interaction through the repulsive spanning tree (red arcs) and the attractive subgraph (black arcs) with $\epsilon_A=0.01$ yields a two-cluster state as shown in Fig.\ \eqref{Picture4}. The members of these clusters are successfully predicted using only the vertex decomposition of the repulsive subgraph as illustrated through the snapshots in subfigures (b-c) of Fig.\ \eqref{Picture4}. Moreover, the two clusters experience a phase difference of $\pi$ as depicted through a snapshot in Fig.\ \eqref{Picture4} (b). The temporal evolution of $y_i$ in Fig.\ \eqref{Picture4} (d) confirms the agents maintain an oscillatory rhythm, and the system undergoes through a two-cluster synchronization. The chosen attractive-repulsive connections portrait that oscillator-$1$ goes through only repulsive interactions, and as a consequence, it will split from the coherent group. %Hence, we find $-{\bf{x}}_1 \approx {\bf{x}}_2={\bf{x}}_3={\bf{x}}_4$ in Fig.\ \eqref{Picture4}.

\subsubsection{Antisynchronization}

\par We pass the repulsive strength $\epsilon_R=-0.1$ through a different spanning tree drawn in Fig.\ \eqref{Picture3} (c). The rest of the chords (the black arcs of Fig.\ \eqref{Picture5} (a)) of $K_4$ are attractively coupled with the coupling strength $\epsilon_A=0.01$. Interestingly, this spanning tree (the red arcs of Fig.\ \eqref{Picture5} (a)) looks like a simple chain. In fact, this repulsive subgraph is a hamiltonian path because it visits each node of $K_4$ exactly once. The set of vertices of the chosen repulsive spanning tree is the union of two disjoint sets $V_1=\{1,3\}$ and $V_2=\{2,4\}$. Again, this vertex decomposition works perfectly well for predicting the members of the clusters, as contemplated in Fig.\ \eqref{Picture5}. The phase difference between these two clusters is exactly $\pi$ (see Fig.\ \eqref{Picture5} (b)) after the initial transient. In addition, the amplitude $r_i$ of each oscillator is identical for all $i=1,2,3,4$. The system \eqref{eq:1} displays two synchronized groups, and the phase difference between these two clusters is $\pi$. In fact, whenever we pass the negative coupling strength through a simple chain $\{1,2,3,\cdots, N\}$ in a complete graph $K_N$ with even $N \geq 2$, we find the amplitude $r_i$ of each oscillator is precisely same. The term chain reflects the graph associated with the symmetric adjacency matrix $A_{i(i+1)}=1$ for $i=1,2,3,\cdots,(N-1)$. Besides, the complete graph $K_N$ with even $N$ for such a repulsive subgraph manifests two clusters, where one group consists of the members with the indices $\{1,3,\cdots, (N-1)\}$ and the other cluster contains the oscillators with the indices $\{2,4,\cdots, N\}$. Furthermore, since the phase difference among the two clusters is $\pi$, thus ${\bf x}_i+{\bf x}_j=\bf{0}$ for $i \in \{1,3,\cdots, (N-1)\}$ and $j \in \{2,4,\cdots, N\}$, which is a common signature of antisynchronization \cite{chowdhury2021antiphase,chowdhury2023interlayer}. During this antisynchronization, if the $i$-th and $k$-th oscilltors are in antisynchronized state, then we have 

\begin{equation} \label{Eq6}
	{\bf x}_i+{\bf x}_k=\bf{0}. 
\end{equation}

%\begin{figure*}[!t]
%	\centerline{\includegraphics[width=0.65\textwidth]{Picture7}}
%	\caption{{\bf Predicting the members of two synchronized clusters  }: (a) Red arcs delineate a spanning tree of the drawn network with $10$ nodes. We pass the positive coupling strength through the black edges. Here, $\epsilon_A=10^{-2}$ and $\epsilon_R=-10^{-1}$. (b) Utilizing the bipartiteness of the used repulsive spanning tree, we can easily predict the oscillators with indices $\{1,3,5,7,9\}$ and $\{2,4,6,8,10\}$ belong to two different clusters. In addition, the phase difference between any two representative oscillators of these respective two clusters is $\pi$. (c) The snapshot at a specific time after the transient validates the occurrence of cluster synchronization. (d) The temporal evolution of $y_i$ attests that the trajectories oscillate in a limit cycle region, maintaining a fixed $\pi$-phase difference between these two synchronized populations. }
%	\label{Picture7}
%\end{figure*}

The time evolution of these two oscillators are given by

\begin{equation} \label{Eq7}
	\begin{split}
		\dot{{\bf x}_i}=f({\bf x}_i)+ \sum_{j=1}^{N} D_{ij} H({\bf x}_i,{\bf x}_j), \text{and}\\
		\dot{{\bf x}_k}=f({\bf x}_k)+ \sum_{j=1}^{N} D_{kj} H({\bf x}_k,{\bf x}_j).
	\end{split}
\end{equation}

Combining the equations \eqref{Eq6} and \eqref{Eq7}, we get

\begin{equation} \label{Eq8}
	\begin{split}
		\dot{{\bf x}_i}=f({\bf x}_i)+ \sum_{j=1}^{N} D_{ij} H({\bf x}_i,{\bf x}_j), \text{and}\\
		\dot{{\bf x}_i}=-f(-{\bf x}_i)- \sum_{j=1}^{N} D_{kj} H(-{\bf x}_i,{\bf x}_j).
	\end{split}
\end{equation}

\par Both of these equations \eqref{Eq8} are compatible and consistent if they satisfy a few conditions. Of which one will definitely be, \textit{$f$ must be an odd function, i.e., $f(-\mathbf{x})=-f(\mathbf{x})$}. Thus, we choose identical SL oscillators on top of each node. Similarly, the other necessary condition for obtaining antisynchronization is \textit{the coupling function $H$ must be an odd function}, and our choice of $H({\bf x}_i,{\bf x}_j)=[{x}_j + {x}_i, 0]^T$ satisfies this condition too. However, such a splitting of two clusters with an equal amplitude for all oscillators can not be anticipated even in a complete network with an odd number of vertices, as we already observed in Fig.\ \eqref{Picture2}. The amplitudes of the two clusters differ by a small magnitude in Fig.\ \eqref{Picture2} (see subfigues (c) and (f)).

%\begin{figure*}[!t]
%	\centerline{\includegraphics[width=0.55\textwidth]{Picture8}}
%	\caption{{\bf Prediction of the members of the clusters through a different spanning tree}: (a) We choose the same network as portrayed in Fig.\ \eqref{Picture7} and pass the negative coupling strength $\epsilon_R=-10^{-1}$ through a different spaning tree contemplated through the red arcs. The attractive coupling strength $\epsilon_A=10^{-2}$ is applied through the black links. The vertex partitioning of the chosen repulsive path provides two disjoint and independent sets $V_1=\{1,4,5,8\}$ and $V_2=\{2,3,6,7,9,10\}$, respectively. (b-c) Subfigures (b) and (c) confirm the members of the two clusters separate as per the vertex partitioning of the repulsive spanning tree. (d) The trajectories in the limit cycle regime sustain coherent behaviour within the same cluster; however, the clusters undergo a phase difference of $\pi$.}
%	\label{Picture8}
%\end{figure*}

\subsubsection{Theoretical Computation of $F$ and $F_{chain}$}

\par Now, we are coming back to the calculation of the frustration of $K_4$ through two different spanning trees shown in Fig.\ \eqref{Picture3}. We plot the respective changes of the frustration indices in Fig.\ \eqref{Picture6}. Since $K_4$ is a non-bipartite graph, we can not expect zero frustration ($F=0$) in this network \cite{chowdhury2020effect}. In both of these chosen repulsive subgraphs, we fix the attractive coupling strength $\epsilon_A=0.01$ and decrease the repulsive coupling strength $\epsilon_R$ with a very small step-length $-0.00001$. Initially, when $\epsilon_R$ is positive beyond a critical value, all the oscillators are in-phase synchronized (in fact, they are oscillating with identical amplitude too). But, as soon as $\epsilon_R$ becomes negative of suitable strength, the asymptotic value of $F$ saturates, as observed in  Fig.\ \eqref{Picture6}. The previous Figs.\ (\ref{Picture4}-\ref{Picture5}) are drawn for $\epsilon_R=-0.1$, for which the values of this frustration $F$ becomes horizontal with respect to the repulsive coupling strength $\epsilon_R$. In both of these subfigures, the red curve illustrates the frustration of the network $K_4$ for the hamiltonian path shown in Fig.\ \eqref{Picture3} (c). The yellow curve represents the same for a different spanning tree shown in Fig.\ \eqref{Picture3} (b). A noticeable observation is that the hamiltonian path of Fig.\ \eqref{Picture3} (c) can provide comparatively less $F$ value than another repulsive spanning tree. Furthermore, our analytical prediction of $F$ for these spanning trees are $\dfrac{2}{3}$ for the hamiltonian path and $1$ for the spanning tree shown in Fig.\ \eqref{Picture3} (b). The red and yellow lines in Fig.\ \eqref{Picture6} (a), i.e., our numerical simulation, exactly agrees with our theoretical prediction. Similarly, the numerical calculations of $F_{chain}$ saturate at the values $0$ (red line), and $1.333$ (yellow line), while our theoretical prediction is $0$ and $\dfrac{4}{3}$, respectively. Another vital inspection of this Fig.\ \eqref{Picture6} is before the asymptotic convergence of both measures $F$ and $F_{chain}$, there exists fluctuations in both of these lines.  Actually, the simulations use random initial conditions from $[-1,1] \times [-1,1]$ for each $\epsilon_R$. Hence, depending on initial conditions, such fluctuations are noticed for both spanning trees. However, beyond a critical value of $\epsilon_R$ for fixed $\epsilon_A$, these values of $F$ and $F_{chain}$ remain unchanged irrespective of any random initial conditions chosen from $[-1,1] \times [-1,1]$.

\subsubsection{The stability of antisynchronous clusters}     	
\par Due to the presence of attractive and repulsive connection schemes, two clusters $\{C_{1}, C_{2}\}$ can be observed, and furthermore, these clusters are in an antisynchronized state for some cases. Let $s_{m}, \; m=1,2$ be the state of synchronous solution of the nodes belonging to the cluster $C_{m}$. Then, the evolution of the synchronous solution is given by, 
\begin{equation} \label{sync_sol}
	\begin{array}{l}
		\dot{s}_{m}= f (s_{m}) + \sum_{j=1}^{2} Q_{mj} H(s_{m},s_{j}), \;\; m=1,2, 
	\end{array}
\end{equation}  
where the $2 \times 2 $ matrix $Q$ represents the quotient network \cite{lodi2021one,panahi2021cluster} associated with the two clusters. 

\par To investigate the stability of the cluster synchronous solution, we consider a small perturbation $\delta \mathbf{x}_{i} = x_{i} - s_{m} $, $i \in C_{m}$. Then the variational equation can be written as, 

\begin{multline} \label{var_eqn}
		\delta \dot{\mathbf{x}}= \bigg[\sum\limits_{m=1}^{2} W^{[m]} \otimes Jf(s_{m})+\sum\limits_{m=1}^{2} (KW^{[m]}) \otimes J_{1}H(s_{m},s_{m'}) \\+ \sum\limits_{m=1}^{2} (DW^{[m]}) \otimes J_{2}H(s_{m},s_{m'}) \bigg] \delta \mathbf{x},\; m'=1,2, 
\end{multline}
where the $dN$-dimensional state vector $ \delta {\mathbf{x}}= \big[\delta {\mathbf{x}}_{1}^{tr}, \delta {\mathbf{x}}_{2}^{tr}, \cdots, \delta {\mathbf{x}}_{N}^{tr}\big]^{tr}$, and $\otimes$ represents the Kronecker product. $W^{[m]}$ is a $N \times N$ diagonal matrix such that, 
\begin{equation}
	\begin{array}{l}
		W^{[m]}_{ii}=\begin{cases*}
			1 , \;\; \mbox{if}  \;\; i \in C_{m} \\
			0, \;\; \mbox{otherwise}. 
		\end{cases*} 
	\end{array}
\end{equation} and $\sum\limits_{m=1}^{2} W^{[m]}= I_{N}$, the $N \times N$ identity matrix. $K$ is the $N\times N$ diagonal matrix whose diagonal entries are given by $K_{ii}=\sum\limits_{j=1}^{N}D_{ij}$, and characterizes the weighted degree of node $i$ of the underlying signed graph. $J_{1}$ and $J_{2}$ stand for the Jacobian operators with respect to the first and second variable, respectively.  
\par Therefore, the problem of stability requires to solve the linearized differential equation \eqref{var_eqn} along with the nonlinear equation \eqref{sync_sol} for maximum Lyapunov exponents. The negative value of maximum Lyapunov exponent transverse to the cluster synchronization manifold gives the necessary condition for stable cluster synchronization state.
\par Now, we can further simplify the variational equation \eqref{var_eqn} when the clusters are in antisynchronized state. The presence of antisynchronization between the two cluster state (i.e., $s_{1} (t) +s_{2} (t) =0$) indicates that the Jacobian matrices $Jf (s_{1})$ and $Jf (s_{2})$ are related to each other, as well as $J_{m}H (s_{j},s_{j'})$ and $J_{m}H(s_{j},s_{j})$, where $m,j,j'=1,2$. Moreover, antisynchronization between the cluster synchronization states require $f(x)$ and $H(x)$ to be odd functions as a necessary condition (Eqs.\ \eqref{Eq6}-\eqref{Eq8}). Combining these two conditions, we can acquire that the Jacobians in Eq.\ \eqref{var_eqn} satisfy the relations $Jf(s_{1}) = Jf(s_{2})$ and $J_{m}H(s_{1},s_{1}) = J_{m}H (s_{2},s_{2})=J_{m}H(s_{1},s_{2})$=$\mathcal{J}_{m}H\; \mbox{(say)}$, for $m=1,2$, since the Jacobian of odd function is even and thus independent of the sign of the variables. This eventually simplifies the variational equation \eqref{var_eqn} as follows,
\begin{multline} \label{var_eqn2}
		\delta \dot{\mathbf{x}}= \bigg[ I_{N}\otimes Jf(s_{1})+ K \otimes \mathcal{J}_{1}H +D \otimes \mathcal{J}_{2}H \bigg] \delta \mathbf{x},
\end{multline}    
where $s_{1}$ denotes the state of nodes in one cluster at coherent state. Therefore, the problem of stability of cluster synchronization state is then reduced to solving the coupled linear differential equation \eqref{var_eqn2} for maximum Lyapunov exponent. For the specific problem where the coupling function $H$ is an odd function of $\mathbf{x}_{i}+\mathbf{x}_{j}$, i.e., $H(\mathbf{x}_{i},\mathbf{x}_{j})=\mathcal{F}(\mathbf{x}_{i}+\mathbf{x}_{j})$, one can further obtain that $\mathcal{J}_{1}H=\mathcal{J}_{2}H=\mathcal{J}H\;\mbox{(say)}$, and thus the stability problem reduced to solving a more simplified variational equation 
\begin{multline} \label{var_eqn3}
	\delta \dot{\mathbf{x}}= \bigg[ I_{N}\otimes Jf(s_{1})+ \mathcal{M} \otimes \mathcal{J}H \bigg] \delta \mathbf{x},
\end{multline}    
where $\mathcal{M}=K+D$.
%\begin{figure*}[!t]
%	\centerline{\includegraphics[width=0.75\textwidth]{Picture11}}
%	\caption{{\bf $F$ and $F_{chain}$ with varying $\epsilon_R$}: Numerically saturated values of $F$ and $F_{chain}$ closely match with our theoretically calculated values of these measures. The Red line represents the results for the repulsive spanning tree in Fig.\ \eqref{Picture10} (b), whereas the yellow line displays the results for the spanning tree in Fig.\ \eqref{Picture10} (c). (a) For both of these non-isomorphic spanning trees, the system saturates to the $F \approx 0.666431904$, while our analytical predicted $F=\dfrac{2}{3}$. (b) Passing the repulsive coupling strength through the spanning tree in Fig.\ \eqref{Picture10} (b), we get $F_{chain} \approx 0$. $F_{chain} \approx 1.42900455$ for the spanning tree in Fig.\ \eqref{Picture10} (c), which is very close to the theoretically predicted $F_{chain}=\dfrac{10}{7}$. Here, $\epsilon_A=10^{-2}$ and $\epsilon_R$ is varied with a small step-length $-10^{-5}$ starting from $10^{-2}$. The initial conditions are drawn randomly from $[-1,1] \times [-1,1]$.}
%	\label{Picture11}
%\end{figure*}

\par To validate the clusters of synchrony in the complete network $K_4$ given in Fig.\ \eqref{Picture5}, we compute the maximum transverse Lyapunov exponent $\Lambda$ (red dashed line) of the Eq.\ \eqref{var_eqn2} and plot the synchronization error $E_{k}$ of the $k$-th cluster (blue line) in Fig.\ \eqref{sw_random}. We define the cluster synchronization error $E_k$ as follows

\begin{equation}\label{error}
	E_{k}=\Bigg\langle \bigg(\frac{1}{Card(C_k)}\sum_{i \in C_k}^{}\left(x_i-\bar{x}_k\right)^2\bigg)^{\frac{1}{2}} \Bigg\rangle_t,
\end{equation}

where $Card(C_k)$ is the cardinality of the set $C_k$, the group of vertices involved in the cluster $k$. $\bar{x}_k$ denotes the average of all vertices within a particular cluster $k$ at time $t$. $\Big\langle \cdots \Big\rangle_t$ represents an average over a sufficiently long time interval. Since the coupled system in Fig.\ \eqref{Picture5} settles down to antisynchronization, both the errors $E_1$ and $E_2$ converge to zero at the same choice of repulsive coupling strength $\epsilon_{R} < 0$. Thus, we plot $E=E_1=E_2$ (blue line) in Fig.\ \eqref{sw_random}. Figure \eqref{sw_random} confirms the local asymptotic stability of cluster synchronization, as the maximum Lyapunov exponent transverse to the cluster synchronization manifold becomes negative almost at the same time where the cluster synchronization error $E$ becomes zero.

\begin{figure*}[htp]
	\centerline{\includegraphics[width=1.0\textwidth]{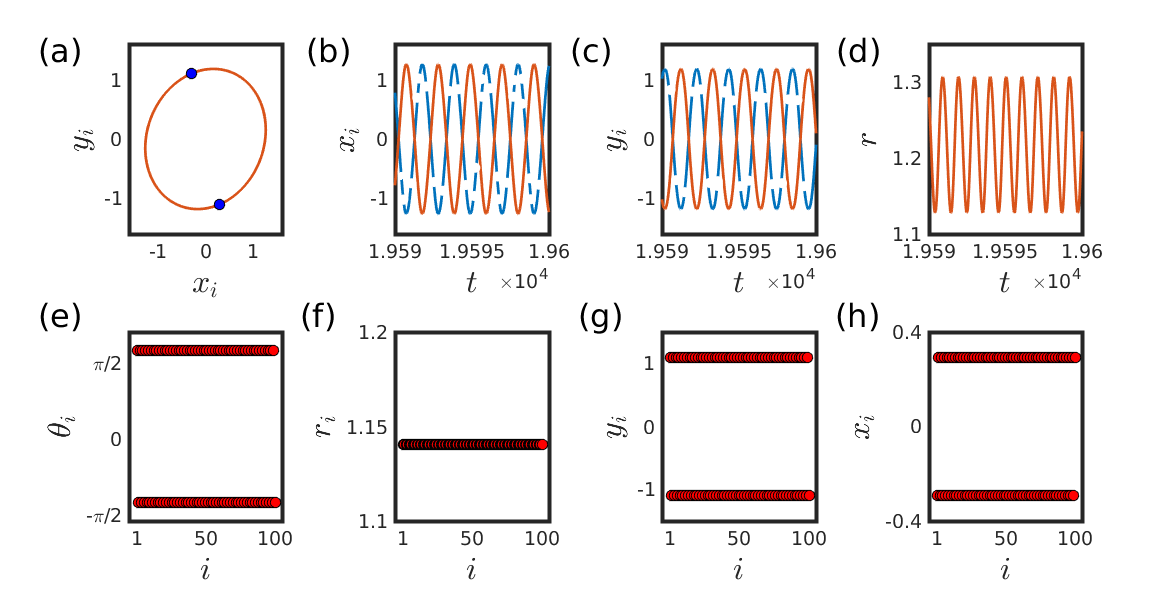}}
	\caption{{\bf Splitting of two distinct clusters in $K_{100}$}: We choose the complete graph $K_{100}$ with $N=100$ vertices and $4950$ edges and pass the repulsive coupling strength $\epsilon_R=-4.0$ through the simple chain containing only $N-1=99$ links. The positive coupling strength is kept fixed at $\epsilon_A=0.01$ and the initial conditions are randomly chosen from $[-1,1] \times [-1,1]$. The phase portrait in subfigure (a) and the temporal evolutions in subfigures (b-c) reveal the emergence of two clusters. The variation of the amplitude $r_i$ of each oscillator is manifested in subfigure (d). (e-h) The lower panel exhibits the snapshots at a particular time after the transients. These snapshots unveil the system settles down a two-cluster state, and the phase difference between these two clusters is $\pi$. Since the amplitude $r_i$ of all oscillators is equal, thus we observe the onset of antisynchronization.}
	\label{Picture12}
\end{figure*}

\begin{figure*}[htp]
	\centerline{\includegraphics[width=1.0\textwidth]{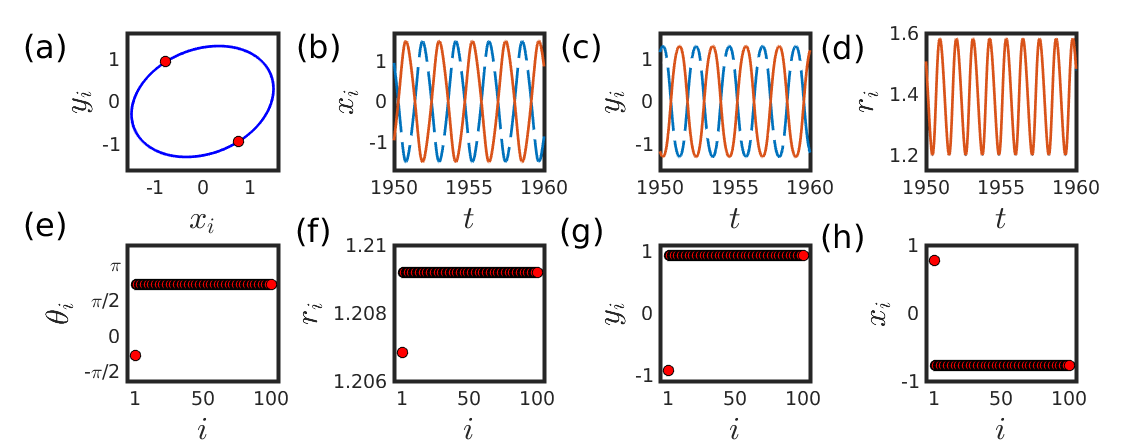}}
	\caption{{\bf The impact of a different spanning tree on $K_{100}$}: (a) The phase portrait in the $xy$-plane, (b) the temporal evolution of $x_i$, and (c) the temporal evolution of $y_i$ ensure the system splits into two synchronized populations. (d) The temporal evolution of $r_i$ is represented in subfigure (d). (e-h) Due to our chosen spanning tree, oscillator-$1$ leaves the other synchronized population, and the phase of this oscillator deviates a distance of $\pi$ from the phase of the other oscillators. The snapshots in the subfigures (e-h) validate our claim of predicting the members of the clusters using the vertex partitioning of the used repulsive spanning tree. Here, $\epsilon_R=-3.6$, and $\epsilon_A=0.01$.}
	\label{Picture13}
\end{figure*}

\begin{figure*}[htp] 
	\centerline{
		\includegraphics[scale=0.2]{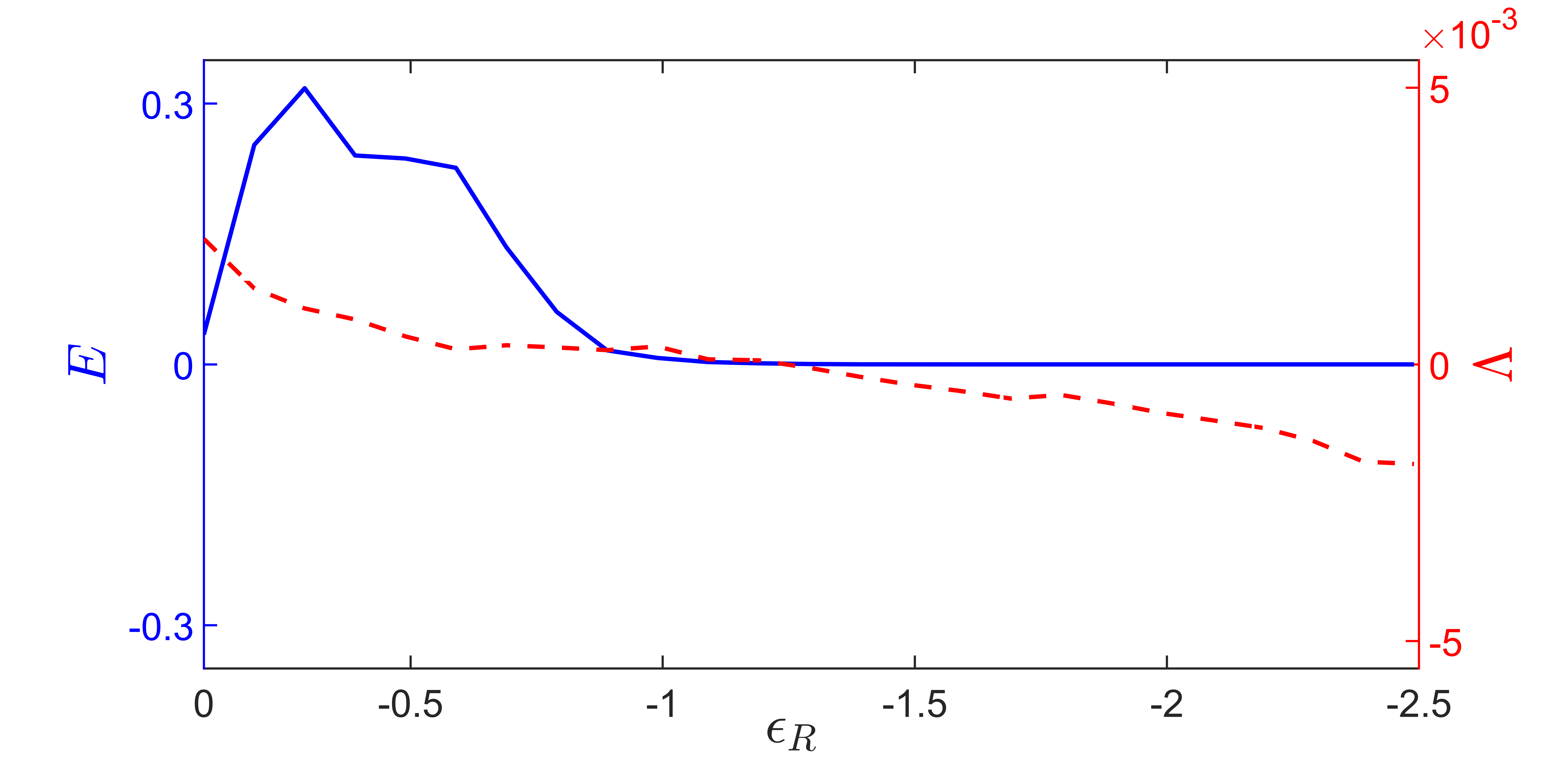}}
	\caption{{\bf The (local) stability of the cluster-synchronized state:} The simulated maximum Lyapunov exponent $\Lambda$ (red dashed line) transverse to the cluster synchronization manifold becomes negative, confirming each cluster's stability. The underlying network $K_{100}$ contains the chain as the repulsive subgraph as illustrated in Fig.\ \eqref{Picture12}. The stability of the synchronous solution is further validated by plotting the cluster synchronization error $E$ (blue line), which drops to zero beyond a specific strength of $\epsilon_{R} < 0$. Here, $\epsilon_{A}=0.01$. Please see the main text for further information. }
	\label{k_global}
\end{figure*}

\subsection{The complete graph $\mathbf{K_{100}}$}

\par Now, we choose again a complete graph which is comparatively larger than other previously considered networks. In Figs.\ (\ref{Picture12}-\ref{Picture13}), we choose the global network with $N=100$ nodes and $L=4950$ links. Cayley's formula {\color{black}predicts} this network contains $100^{98}$ number of spanning trees. To represent our findings, we select two different non-isomorphic spanning trees. Each spanning tree contains exactly $(N-1)=99$ links. We pass the negative coupling strength $\epsilon_R$ of adequate strength through these spanning trees. Such a minimum fraction of repulsive edges is able to give rise to two distinct clusters, and these two clusters maintain a phase difference of $\pi$. For Fig.\ \eqref{Picture12}, we pass the repulsive strength $\epsilon_R=-4.0$ through a chain $1-2-3-\cdots-98-99-100$. Since we increase the network size significantly, thus we have to enhance the magnitude of the repulsive strength $\epsilon_R$ too, although we keep the attractive coupling $\epsilon_A=0.01$ as earlier. The result remains unchanged for random initial conditions from $[-1,1] \times [-1,1]$. As discussed earlier, such a choice of repulsive coupling through the chain within attractively coupled SL oscillators in any complete network with an even number of vertices can induce antisynchronization. Our chosen repulsive chain in $K_{100}$ also produces the same. Figure \eqref{Picture12} portrays that the identical SL oscillators break into two halves $V_1=\{1,3,5,\cdots,99\}$ and $V_2=\{2,4,6,\cdots,100\}$ analogous to the vertex partitioning to our selected spanning tree. These two synchronized clusters oscillate opposite in phase with each other. Since the amplitude $r_i$ of each oscillator is identical, hence ${\bf x}_i+{\bf x}_k=\bf{0}$ for $i \in V_1$ and $k \in V_2$. One needs to iterate the system for sufficiently long to achieve the antisynchronization in $K_{100}$ with the repulsive chain.

\par We choose a different spanning tree with the set of vertices $V_1 \cup V_2$ in Fig.\ \eqref{Picture13}, where $V_1=\{1\}$ and $V_2=\{2,3,4,\cdots,98,99,100\}$. Since the oscillator-$1$ experiences only repulsive interactions with other $N-1=99$ oscillators as per the arrangement of our spanning tree. Hence, this oscillator-$1$  leaves the coherent group, and the system manifests the occurrence of a solitary state. The emergence of such a solitary state is also contemplated in Figs.\ (\ref{Picture2}) and (\ref{Picture4}). The adjacency matrix $C$ corresponding to the chosen spanning tree is given by 

\[ \begin{cases} 
	C_{1j}=C_{j1}=1 & \text{for} \hspace{0.2cm} j=2,3,4,\cdots,N \\
	\hspace{0.9cm}0 &  \text{otherwise}
\end{cases}
\]

Hence, the equation \eqref{eq:1} yields

\begin{equation}
	\begin{split}
		\dot{{\bf x}_1}=f({\bf x}_1)+ \epsilon_R\sum_{j=2}^{N} H({\bf x}_1,{\bf x}_j), \hspace{0.2cm}\\
		\dot{{\bf x}_i}=f({\bf x}_i)+ \epsilon_{A}\sum_{j=1}^{N} B_{ij} H({\bf x}_i,{\bf x}_j)+\epsilon_R H({\bf x}_i,{\bf x}_1),\\i=2,3,4,\cdots,N
	\end{split}
\end{equation}

Due to the dissimilarity between these two equations, the temporal evolution of oscillator-$1$ does not coincide with the other synchronized cluster with members of $V_2=\{2,3,4,\cdots,N\}$. This understanding is consistent with the appearance of solitary state in   Figs.\ \eqref{Picture4}, and \eqref{Picture13}, respectively. 
To validate the stability of each cluster, we compute the maximum transverse Lyapunov exponent $\Lambda$ (red dashed line) of the Eq.\ \eqref{var_eqn2} and plot the cluster synchronization error $E=E_{1}=E_2$ (blue line) of the Eq.\ \eqref{error} in Fig.\ \eqref{k_global}. The negativity of $\Lambda$ ensures the damping of the transversal modes and assures the (local) stability of the cluster-synchronized manifold. Moreover, the cluster synchronization error $E$ descends to zero and leads to the convergence towards the cluster synchronized state.

\par For Fig.\ \eqref{Picture13}, we consider $\epsilon_R=-3.6$ and $\epsilon_A=0.01$, while for Fig.\ \eqref{Picture12}, we consider $\epsilon_R=-4.0$ with same $\epsilon_A$. The choice of the spanning tree in Fig.\ \eqref{Picture13} produces an unbounded solution for $\epsilon_R=-4.0$ and $\epsilon_A=0.01$. Hence, we restrict the negative coupling strength at $\epsilon_R=-3.6$ for fixed $\epsilon_A=0.01$. Figure \eqref{Picture13} confirms the cluster synchronization in the signed network. The phase difference between these two clusters is $\pi$. {\color{black}In Fig.\ \eqref{k_global}, it is evident that the cluster-synchronized state remains locally stable even with a lower repulsive coupling strength. Hence, one can expect to observe similar behavior even at $\epsilon_R=-3.6$.  Thus, while it is possible to set the repulsive coupling strength to $-3.6$ and compare the outcomes in Figs. \eqref{Picture12} and \eqref{Picture13}, our intention was to convey the message that beyond a critical value of the repulsive coupling strength, the solution may become unbounded. As observed, one spanning tree may permit choosing the negative coupling strength up to $-4$, while another may restrict going beyond $-3.6$. To clarify further, our aim is to emphasize the generality of the results: when passing the repulsive coupling of suitable strength through any one of the spanning trees of a connected network, it will split into two coherent clusters with a $\pi$ phase difference. However, the range of the effective repulsive strength may vary depending on the choice of the spanning trees in the connected network.}

\begin{figure*}[!t]
	\centerline{\includegraphics[width=0.65\textwidth]{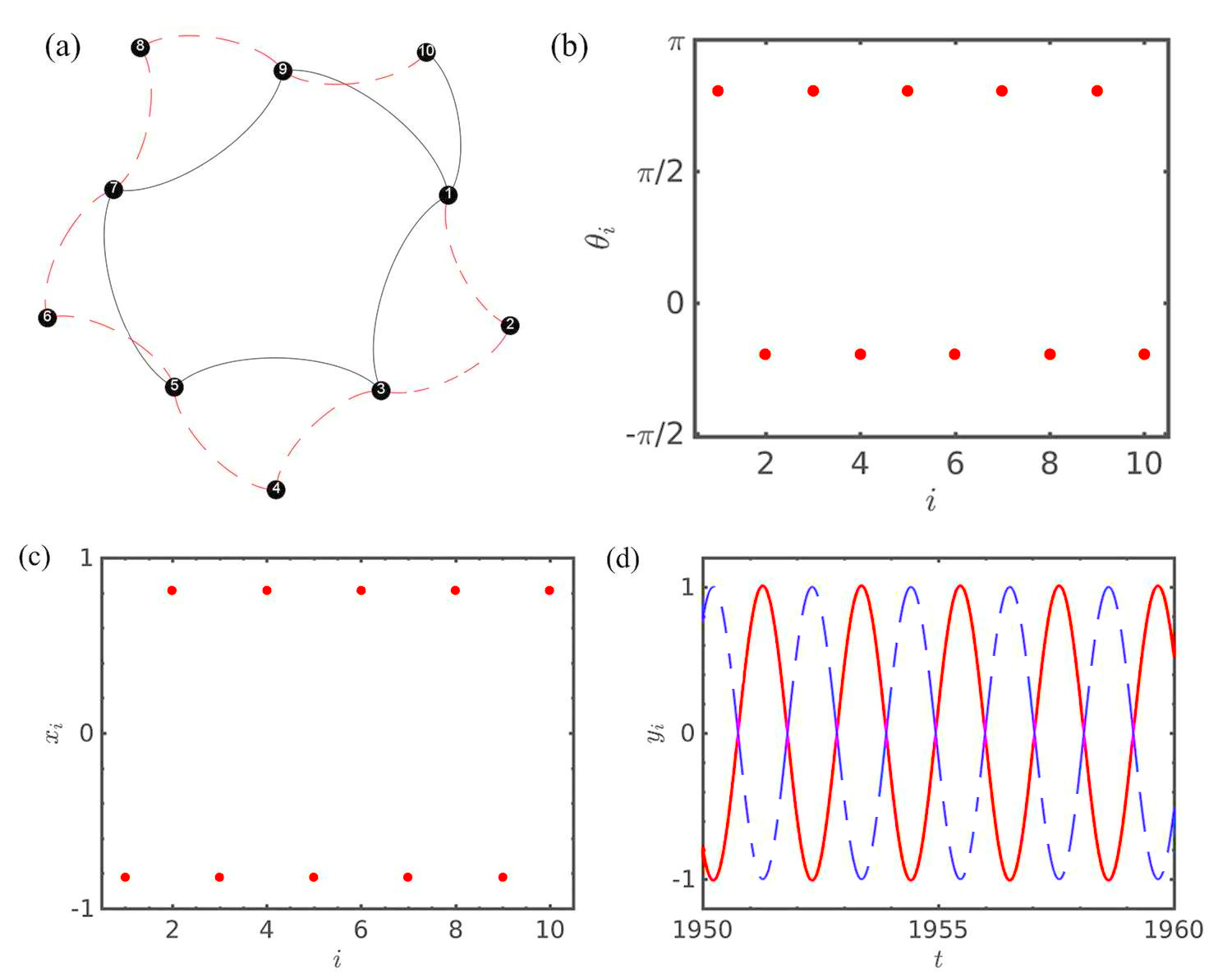}}
	\caption{{\bf Predicting the members of two synchronized clusters  }: (a) Red dashed arcs delineate a spanning tree of the drawn network with $10$ nodes. We pass the positive coupling strength through the black edges. Here, $\epsilon_A=10^{-2}$ and $\epsilon_R=-10^{-1}$. (b) Utilizing the bipartiteness of the used repulsive spanning tree, we can easily predict the oscillators with indices $\{1,3,5,7,9\}$ and $\{2,4,6,8,10\}$ belong to two different clusters. In addition, the phase difference between any two representative oscillators of these respective two clusters is $\pi$. (c) The snapshot at a specific time after the transient validates the occurrence of cluster synchronization. (d) The temporal evolution of $y_i$ attests that the trajectories oscillate in a limit cycle region, maintaining a fixed $\pi$-phase difference between these two synchronized populations. }
	\label{Picture7}
\end{figure*}

\begin{figure*}[!t]
	\centerline{\includegraphics[width=0.65\textwidth]{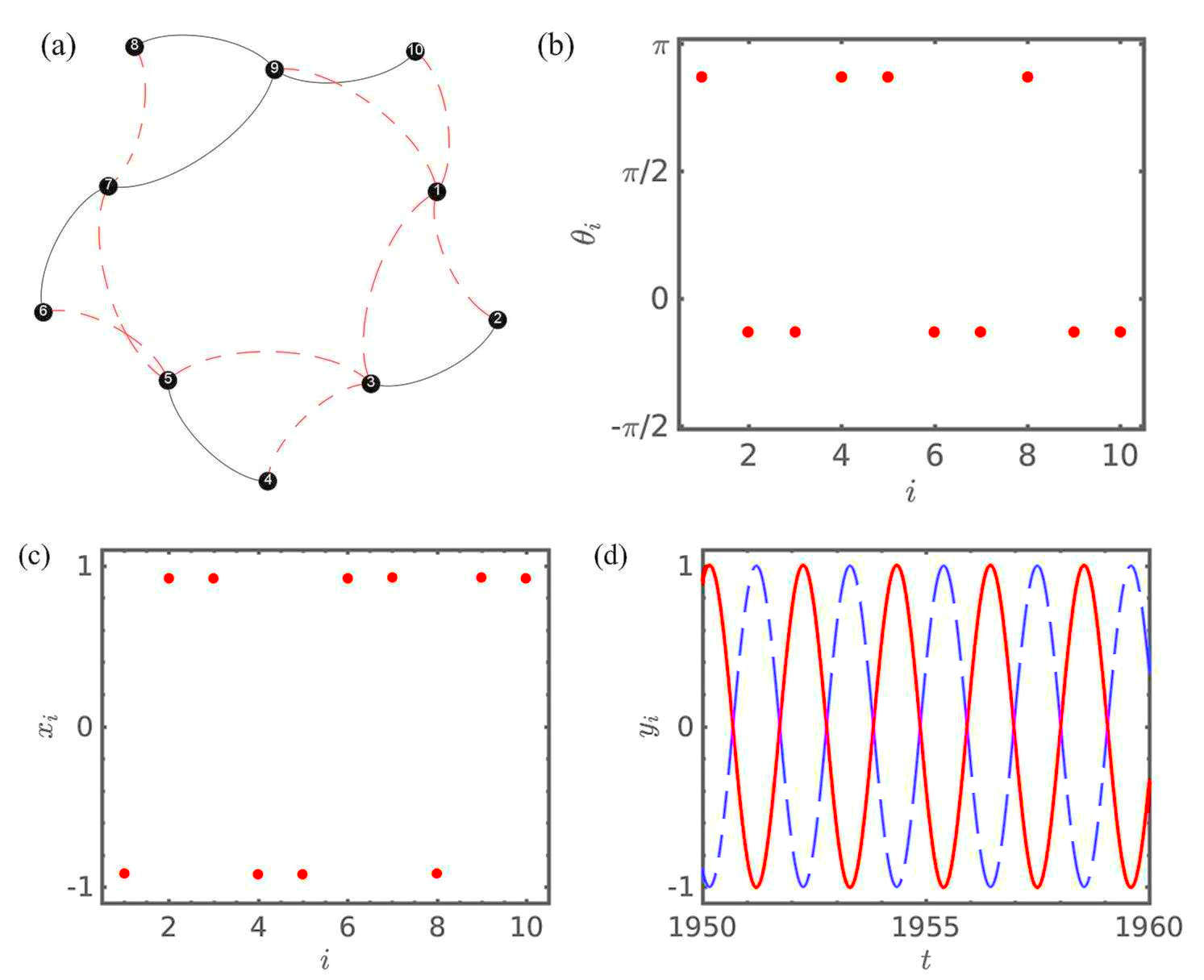}}
	\caption{{\bf Prediction of the members of the clusters through a different spanning tree}: (a) We choose the same network as portrayed in Fig.\ \eqref{Picture7} and pass the negative coupling strength $\epsilon_R=-10^{-1}$ through a different spaning tree contemplated through the red dashed arcs. The attractive coupling strength $\epsilon_A=10^{-2}$ is applied through the black links. The vertex partitioning of the chosen repulsive path provides two disjoint and independent sets $V_1=\{1,4,5,8\}$ and $V_2=\{2,3,6,7,9,10\}$, respectively. (b-c) Subfigures (b) and (c) confirm the members of the two clusters separate as per the vertex partitioning of the repulsive spanning tree. (d) The trajectories in the limit cycle regime sustain coherent behaviour within the same cluster; however, the clusters undergo a phase difference of $\pi$.}
	\label{Picture8}
\end{figure*}

\section{Conclusion} \label{Conclusion}

\par The key finding of this article is the repulsive information propagation of an adequate strength through the existing spanning tree(s) of a connected network can easily induce two distinct groups. The oscillators within the same group are in a coherent state,
%The oscillators within the same group lie within a closed ball of a small radius,
and oscillators of different groups manifest a phase difference of $\pi$. Such splitting of two synchronized clusters is thus possible in any connected graph, as each connected network possesses at least one spanning tree. Moreover, we can successfully anticipate the members of each cluster utilizing the vertex partitioning of the repulsive spanning tree. We successfully analyze the local stability of the cluster synchronous solution by adopting the master stability function (MSF) approach \cite{pecora1998master,anwar2022stability}. In addition, we analyze that

\begin{itemize}
	\item There exists at least one spanning tree in every globally connected network so that whenever we pass the repulsive coupling through this spanning tree, the signed networks with competitive interactions generate a solitary state.
	\item The repulsive coupling introduced through the simple chain can originate antisynchronization in any complete graph with even vertices. For the appearance of antisynchronization, the isolated oscillators dynamics given by $f$ must be an odd function. Also the coupling function $H$ must be an odd function.  
\end{itemize}

The opposite interaction may create frustration in the coupled system. We are able to predict this frustration too in advance (before numerical simulations). Besides, we also calculate the frustration of the chain. Whenever this $F_{chain}$ is equal to zero, each $i$-th oscillator maintains $\pi$-phase difference with the $(i+1)$-th oscillator for $i=1,2,,3,\cdots,(N-1)$. Thus, the system gives rise to an exciting formation, where the vertices with odd indices lie in one group, and other nodes with even indices belong to a different cluster maintaining $\pi$ phase difference. We believe all these findings will foster our understanding of the peculiar behavior of systems with both positive-negative mixed couplings.

%\section*{AUTHOR DECLARATIONS}

\subsection*{ACKNOWLEDGEMENTS}

We thank Francesco Sorrentino for discussions and comments. SNC was supported by NSF grant DMS-1840221.

\begin{figure*}[!t]
	\centerline{\includegraphics[width=0.65\textwidth]{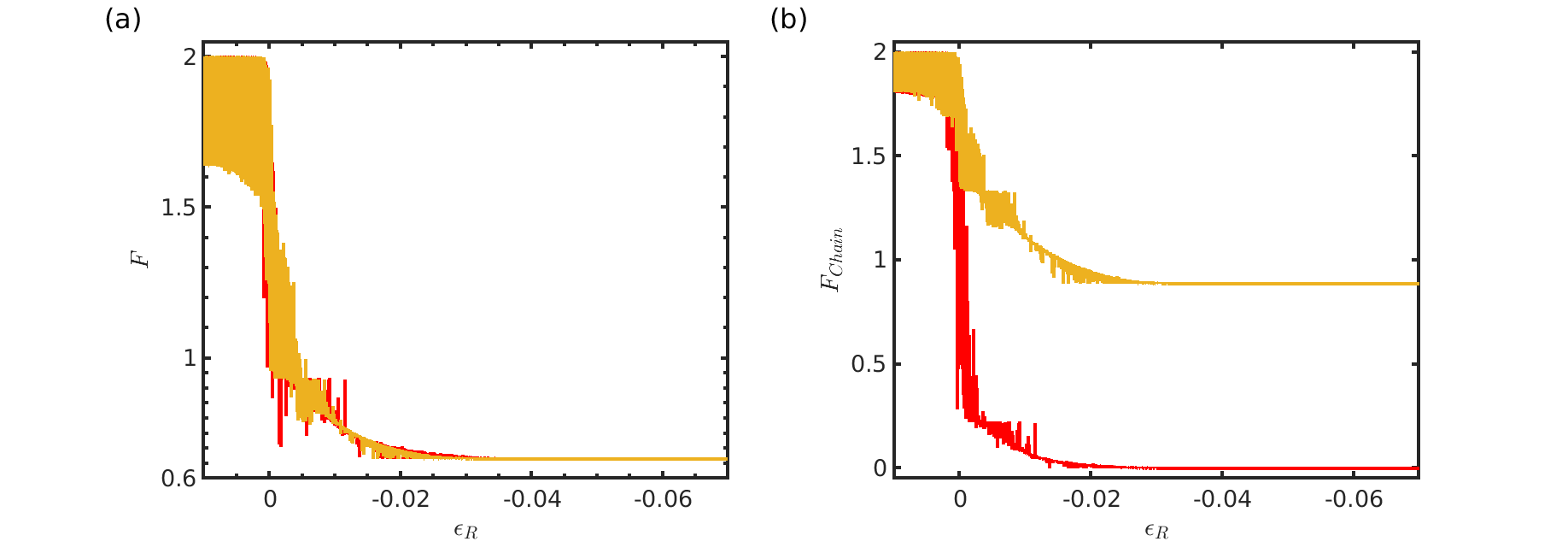}}
	\caption{{\bf Numerically calculated frustration indices $F$ and $F_{chain}$ for two non-isomorphic spanning trees}: (a) Despite the chosen two spanning trees being different as depicted in Figs.\ \eqref{Picture7} (a) and \eqref{Picture8} (a), still, the emergent frustration in the whole network remains the same in the saturated domain. (b) However, the asymptotically saturated frustration that arises in the subgraph labeled as the chain in the article is zero (red) for the spanning tree shown in Fig.\ \eqref{Picture7} (a) and $0.889206886$ (yellow) for the spanning tree used in Fig.\ \eqref{Picture8} (a). We vary the negative coupling strength $\epsilon_R$ from $10^{-2}$ with a tiny step size of $-10^{-5}$ in both these subfigures by keeping fixed the attractive coupling strength at $\epsilon_A=0.01$. Before the saturation, we observe random fluctuations in the values of these two measures $F$ and $F_{chain}$ due to the choice of random initial conditions from $[-1,1] \times [-1,1]$ at each step. {\color{black}Every data point in this figure represents a single realization of the process.}}
	\label{Picture9}
\end{figure*}

\begin{figure*}[!t]
	\centerline{\includegraphics[width=0.65\textwidth]{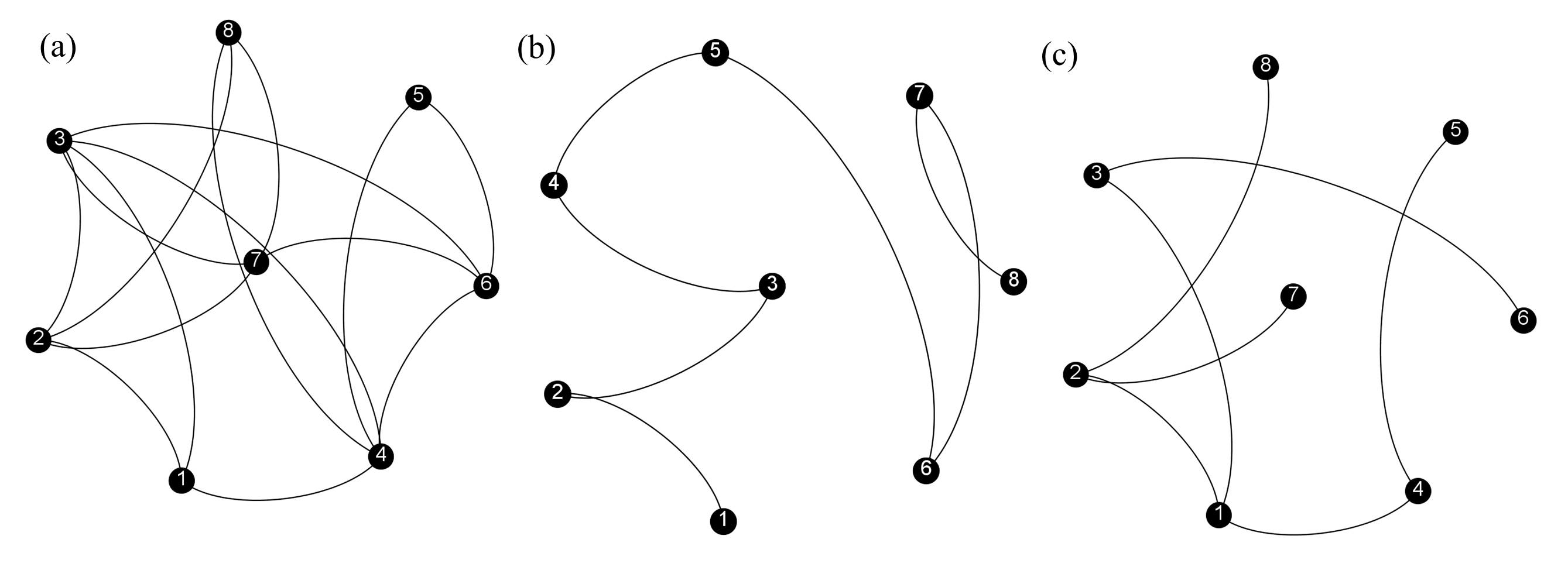}}
	\caption{{\bf A randomly chosen graph along with its two different spanning trees}: (a) A non-bipartite network with $8$ nodes and $15$ links is drawn here. (b-c) We trace this network's two distinct non-isomorphic spanning trees as shown in subfigures (b-c). The vertex partitioning of the spanning tree in subfigure (b) is $V_1=\{1,3,5,7\}$ and $V_2=\{2,4,6,8\}$. Whereas the set of vertices for the spanning tree in subfigure (c) is $V_1=\{1,5,6,7,8\}$ and $V_2=\{2,3,4\}$.}
	\label{Picture10}
\end{figure*}

\begin{figure*}[!t]
	\centerline{\includegraphics[width=0.65\textwidth]{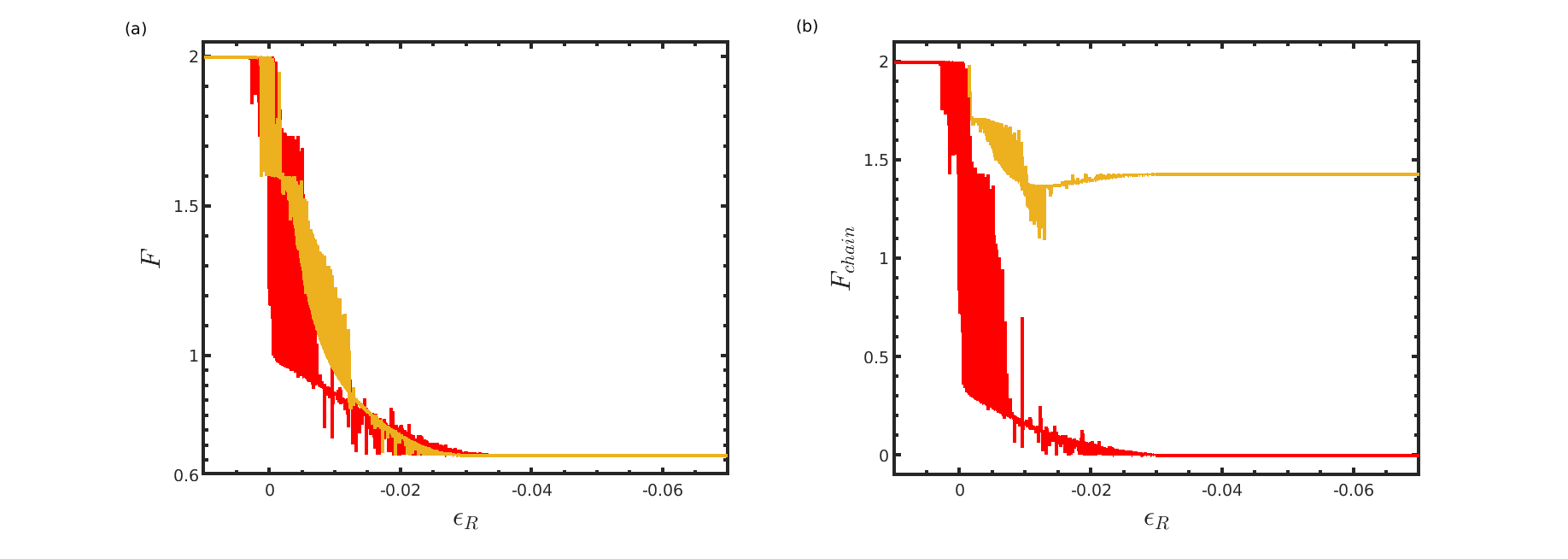}}
	\caption{{\bf $F$ and $F_{chain}$ with varying $\epsilon_R$}: Numerically saturated values of $F$ and $F_{chain}$ closely match with our theoretically calculated values of these measures. The Red (dark gray) line represents the results for the repulsive spanning tree in Fig.\ \eqref{Picture10} (b), whereas the yellow (light gray) line displays the results for the spanning tree in Fig.\ \eqref{Picture10} (c). (a) For both of these non-isomorphic spanning trees, the system saturates to the $F \approx 0.666431904$, while our analytical predicted $F=\dfrac{2}{3}$. (b) Passing the repulsive coupling strength through the spanning tree in Fig.\ \eqref{Picture10} (b), we get $F_{chain} \approx 0$. $F_{chain} \approx 1.42900455$ for the spanning tree in Fig.\ \eqref{Picture10} (c), which is very close to the theoretically predicted $F_{chain}=\dfrac{10}{7}$. Here, $\epsilon_A=10^{-2}$ and $\epsilon_R$ is varied with a small step-length $-10^{-5}$ starting from $10^{-2}$. The initial conditions are drawn randomly from $[-1,1] \times [-1,1]$. {\color{black}Every data point in this figure represents a single realization of the process.}}
	\label{Picture11}
\end{figure*}

\appendix
\section*{Appendix}

In our main text, we have explored the outcomes concerning complete networks. Here, we present results about random networks. These findings illustrate the broad applicability of our discoveries to a wide range of connected networks. {\color{black} We also allocate two subsections for detailed exploration. The fifth subsection outlines the enumeration of various spanning trees within the complete graph $K_4$. In the sixth subsection, we conduct a microscopic examination of the forward and backward transitions of $F$, varying the coupling strength $\epsilon_R$ specifically for the selected spanning tree.}

\subsection{A network with $\mathbf{10}$ nodes} \label{10_node}

\begin{figure*}[!t]
	\centerline{\includegraphics[width=0.65\textwidth]{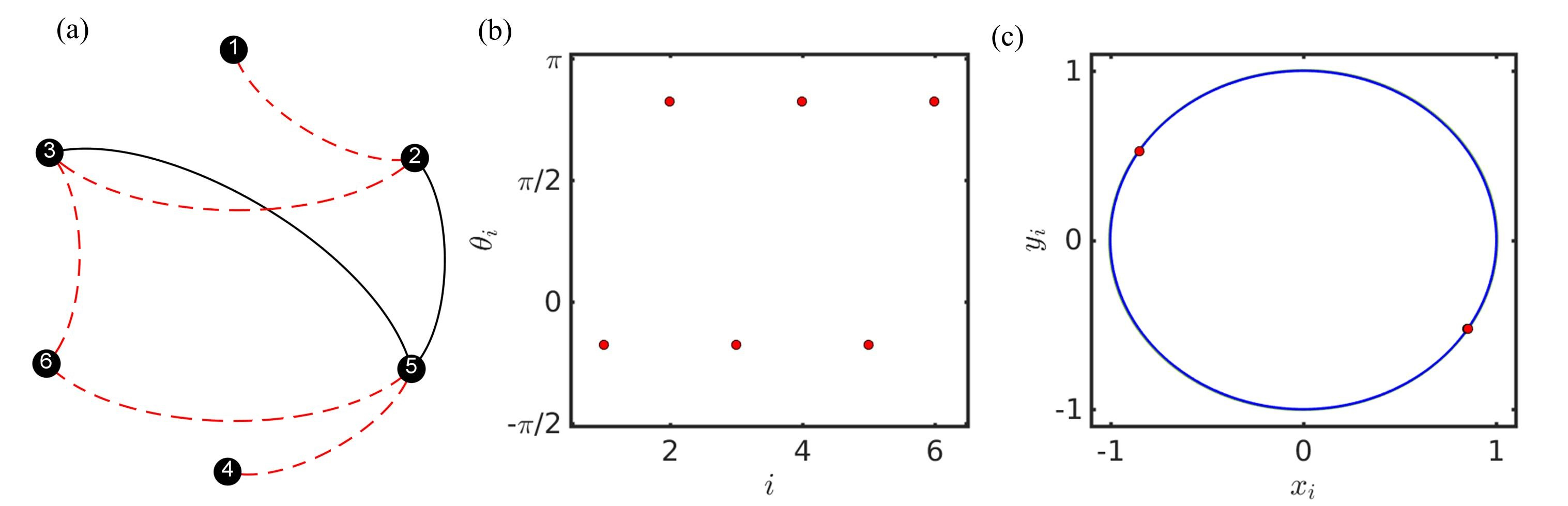}}
	\caption{{\bf Cluster synchronization in a network of $N=6$ non-bipartite network}: (a) We choose a non-bipartite network that does not contain the chain. Still, we trace a spanning tree whose bipartiteness provides two disjoint sets $V_1=\{1,3,5\}$ and $V_2=\{2,4,6\}$.  We pass the repulsive coupling strength $\epsilon_R=-0.1$ through the red dashed arcs and the attractive coupling strength $\epsilon_A=0.01$ through the black links. (b) If we choose one oscillator from the set $V_1$ and another one from the set $V_2$, then the phase difference between that two oscillators is $\pi$ as described through this snapshot. (c) The system splits into two groups, as revealed through the evolution of the oscillators in the $xy$-plane.}
	\label{Picture14}
\end{figure*}

\par Still, now, we present results on complete graphs only. But our prediction algorithm through the vertex decomposition of the repulsive spanning tree works quite well even in other connected networks too. For instance, we choose a network with $N=10$ nodes and $L=15$ edges in Fig.\ \eqref{Picture7}. Using Kirchhoff's matrix-tree theorem \cite{kocay2016graphs}, we calculate this network contains $810$ spanning trees. We select only two non-isomorphic spanning trees. The first one is a simple chain of $N=10$ nodes connecting the $i$-th and $(i+1)$-th oscillators for $i=1,2,3,\cdots,(N-1)$. The repulsive edges, i.e., the branches, are highlighted using red arcs, and the chords, i.e., the attractive links, are represented by black arcs in the subfigure (a) of Fig.\ \eqref{Picture7}. The oscillators rearrange themselves analogous to the vertex partitioning of the chosen spanning tree irrespective of the selected initial conditions from $[-1,1] \times [-1,1]$. Figure \eqref{Picture7} (b) reflects that the phase difference of the oscillators within the same group is nearly zero; nevertheless, the phase difference between the two clusters is $\pi$ under numerical accuracy. Figure \eqref{Picture7} (c) demonstrates the onset of cluster synchronization through a snapshot at a particular time-step after the initial transient. Figure \eqref{Picture7} (d) reveals the oscillatory rhythm is maintained despite the presence of mixed interactions.

\par Figure \eqref{Picture8} demonstrates the consistency of our findings. In this figure, we use the same non-bipartite network considered in Fig.\ \eqref{Picture7} (a) but with a different spanning tree. This considered repulsive spanning tree in Fig.\ \eqref{Picture8} (a) is non-isomorphic to the earlier contemplated spanning tree in Fig.\ \eqref{Picture7} (a). Again, the identical SL oscillators settle down to a two-cluster state, and the members of each cluster can be easily identified using the vertex partitioning of the repulsive spanning tree. Figure \eqref{Picture8} (b) confirms in addition to the spontaneous formation of cluster synchronization, the clusters obey a $\pi$ phase difference between them. The oscillators from the sets $V_1=\{1,4,5,8\}$ and $V_2=\{2,3,6,7,9,10\}$ represent two different groups as shown in Fig.\ \eqref{Picture8} (c). The temporal evolution of these two groups are shown in Fig.\ \eqref{Picture8} (d).

\par Both the Figs.\ (\ref{Picture7}-\ref{Picture8}) are drawn for fixed $\epsilon_A=0.01$ and $\epsilon_R=-0.1$. This repulsive coupling strength $\epsilon_R$ is chosen from the saturated domain of $F$, as depicted through Fig.\ \eqref{Picture9}. yellow lines in Fig.\ \eqref{Picture9} delineate the result for the repulsive spanning tree used in Fig.\ \eqref{Picture8}. The theoretical prediction of $F$ and $F_{chain}$ are $\dfrac{2}{3}$ and $\dfrac{8}{9}$ respectively, which are perfectly match with our numerically found $F \approx 0.666$ and $F_{chain} \approx 0.889$ for this spanning tree. The red one in Fig.\ \eqref{Picture9} is the numerically evaluated frustrations for the repulsive spanning tree shown in Fig.\ \eqref{Picture7} (a).  For this repulsive subgraph, $F$ and $F_{chain}$ saturates to the values $0.666$ and $0$, respectively, which reveals an excellent agreement with our theoretical prediction $F=\dfrac{2}{3}$, and $F_{chain}=0$. Note that both the non-isomorphic spanning trees lead to the same $F$ values for this network. The initial fluctuations in the numerical simulations are due to the choice of random initial conditions from $[-1,1] \times [-1,1]$ for each $\epsilon_R$. Our predicted values of frustrations agree with the numerical simulations in the saturated domain of $F$.

\begin{figure*}[!t]
	\centerline{\includegraphics[width=0.65\textwidth]{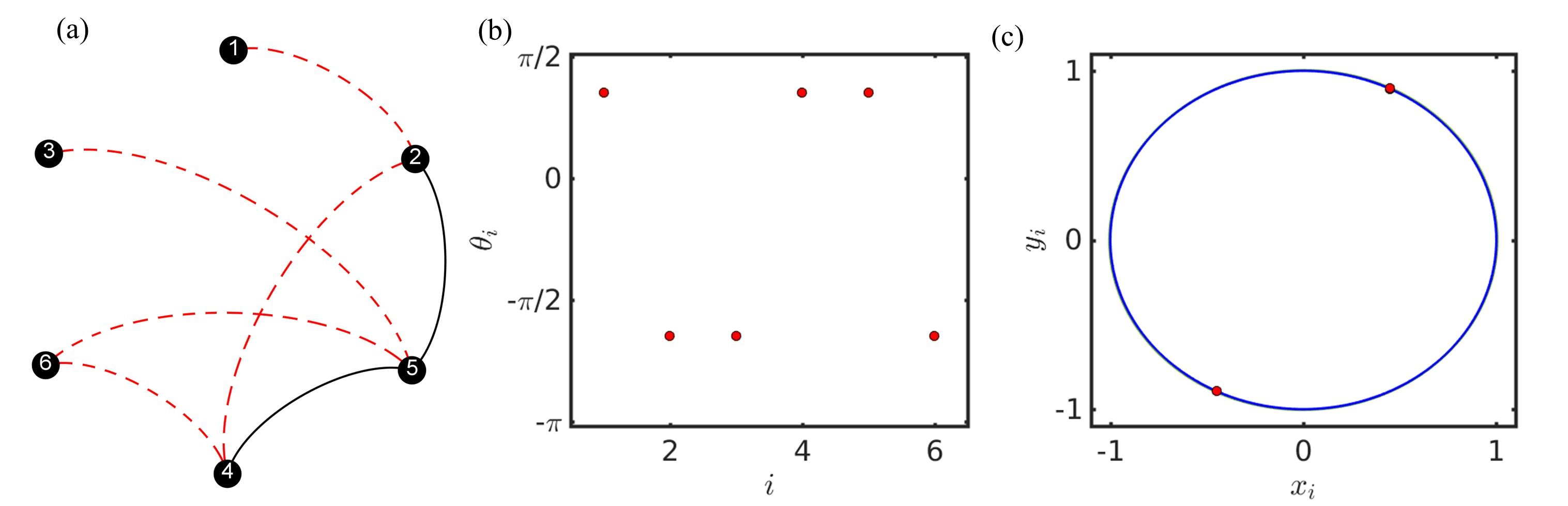}}
	\caption{{\bf Predicting the members of the clusters through the vertex partitioning of the repulsive spanning tree}: (a) The set of vertices of the chosen spanning tree (red dashed arcs) can be split into two disjoint sets $V_1=\{1,4,5\}$ and $V_2=\{2,3,6\}$. Here, $\epsilon_A=0.01$ is passed through the black links of the chosen network of $N=6$ vertices. (b) The vertex decomposition of the repulsive bipartite subgraph allows predicting the two clusters appropriately. The snapshot verifies the $\pi$-phase difference between those two clusters. (c) The formation of two clusters separated by a phase difference of $\pi$ is portrayed here. } 
	\label{Picture15}
\end{figure*}

\subsection{A network with $\mathbf{8}$ nodes} \label{8_node}

We choose a different network with $N=8$ nodes and $L=15$ links (see Fig.\ \eqref{Picture10} (a)) for further numerical assessment. Kirchhoff's matrix-tree theorem provides the number of spanning trees for this network is $1606$. We provide the validation of our findings here by randomly selecting two non-isomorphic spanning trees as presented in Fig.\ \eqref{Picture10} (b-c). Again, we find the splitting of two clusters for both of these repulsive spanning trees (the results are not shown here to avoid the monotonicity). We fix the coupling strengths at $\epsilon_A=0.01$ and $\epsilon_R=-0.1$. The SL oscillators split into two groups maintaining the vertex decomposition of the chosen repulsive spanning trees. These two synchronized groups exhibit a $\pi$ phase difference. All the figures of networks are drawn in this article using the software Gephi \cite{bastian2009gephi}.

\par These selected spanning trees can induce the same amount of frustration in the whole network. To calculate $F$ for both of these repulsive subgraphs, we plot the variation of $F$ in Fig.\ \eqref{Picture11} (a) by changing the coupling strength $\epsilon_R$. The positive coupling strength $\epsilon_A=0.01$ is kept fixed. When $\epsilon_R$ is positive, then the oscillators are almost in-phase synchronized. However, beyond a critical value of $\epsilon_R$, the values of $F$ saturate to $0.666$ for both of the chosen spanning trees. This numerically saturated value of $F$ agrees well with our predicted $F=\dfrac{2}{3}$. For the repulsive spanning tree in Fig.\ \eqref{Picture10} (b), the oscillators split into two groups $V_1=\{1,3,5,7\}$ and $V_2=\{2,4,6,8\}$. Hence, $F_{chain}$ should be zero for this repulsive chain as per our theoretical analysis. This expectation is confirmed in Fig.\ \eqref{Picture11} (b) through the asymptotic convergence of $F_{chain}$ to zero (see the red line in Fig.\ \eqref{Picture11} (b)). The spanning tree in Fig.\ \eqref{Picture10} (c) creates two different clusters. Each of these groups is given by $V_1=\{1,5,6,7,8\}$ and $V_2=\{2,3,4\}$. The presence of links $2-3$, $3-4$, $5-6$, $6-7$, and $7-8$ within the same set contributes to $F_{chain}$. And thus the predicted value of $F_{chain}$ is $\dfrac{2 \times 5}{7}=\dfrac{10}{7}$. Our prediction works quite well in the saturated domain (see the yellow line in Fig.\ \eqref{Picture11} (b)). The value of $\epsilon_R=-0.1$, for which we notice the emergence of two clusters with $\pi$ phase difference, is chosen from the saturated domain as reflected through Fig.\ \eqref{Picture11}. Once the value of $F$ converges, it will not fluctuate further irrespective of the choice of initial conditions within $[-1,1] \times [-1,1]$. A similar argument is valid for the measure $F_{chain}$, too.

\begin{figure*}[htp]
\centerline{\includegraphics[width=1.00\textwidth]{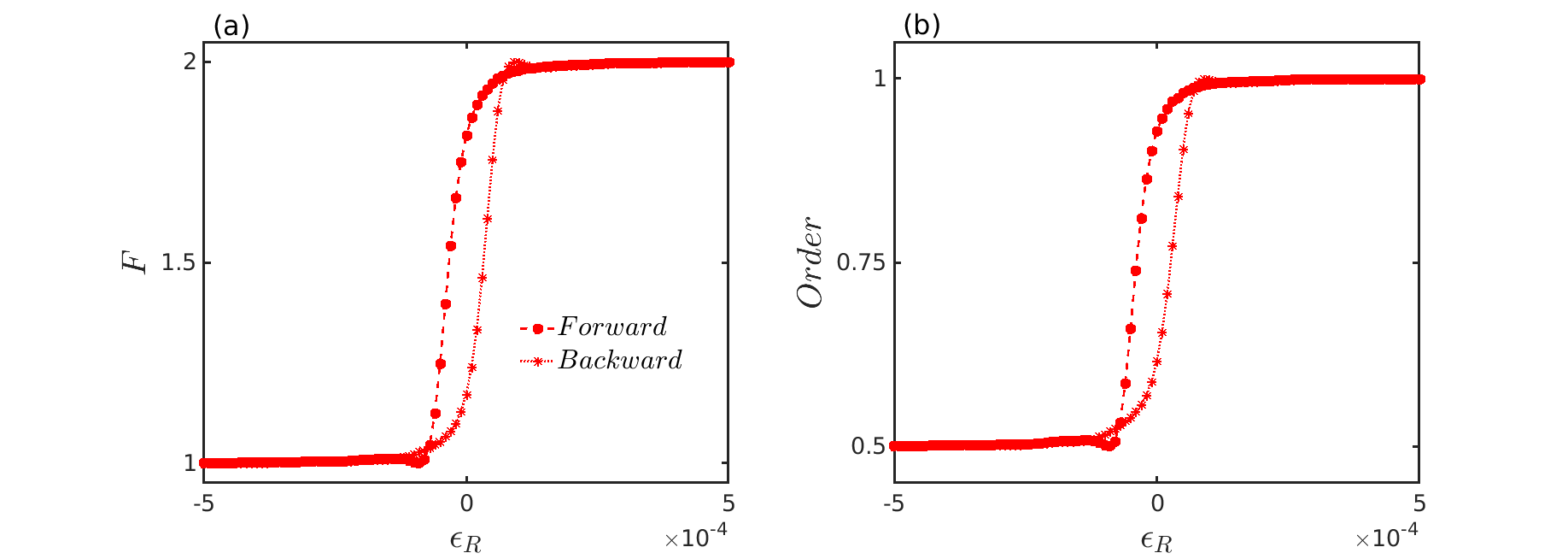}}
\caption{{\bf Variation of $F$ and $Order$ with respect to $\epsilon_R$}: {\color{black}We systematically vary the coupling strength $\epsilon_R$ with a fixed step length of $0.00001$. For the forward transition, $\epsilon_R$ ranges from $0.01$ to $-0.03$; conversely, for the backward transition, it ranges from $-0.03$ to $0.01$. Initially, all conditions are randomly chosen from $[-1,1] \times [-1,1]$. Subsequently, the last point from each simulation iteration serves as the initial conditions for the next iteration, with a small perturbation added to each oscillator from the interval $[0,0.01]$. Notably, the coupling strength $\epsilon_R$ marking the forward transition ($F=2$ to $F<2$) differs from that of the backward transition ($F=1$ to $F>1$). This distinction is evident in the subfigure, depicting the variation of the Kuramoto order parameter, denoted as $Order$. Despite multiple repetitions with different initial conditions, the observed transition remains consistent. Every data point in this figure represents a single realization of the process.}}
\label{Picture105}
\end{figure*}

\subsection{A network with $\mathbf{6}$ nodes without containing the chain as a subgraph} \label{6_node}

\par Till now, we have considered all networks containing the chain $1-2-3-\cdots-(N-2)-(N-1)-N$ as a subgraph. Suppose that chain is not a part of the graph. Then \textit{can we anticipate similar results for such networks?} To answer this query, we select a graph of $N=6$ vertices and $L=7$ links in Fig.\ \eqref{Picture14} (a). This network contains $8$ different spanning trees and does not contain the chain as its subgraph. A spanning tree $1-2-3-6-5-4$ is highlighted through the red arcs. The vertex partitioning of this spanning tree yields two disjoint sets $V_1=\{1,3,5\}$ and $V_2=\{2,4,6\}$. Interestingly, the subgraph termed as `chain' in this study contains the same vertex decomposition.  The coupled oscillators under the influence of this repulsive spanning tree give rise to two cluster states, as shown in Figs.\ \eqref{Picture14} (b-c). The black arcs in Fig.\ \eqref{Picture14} (a) represent the attractive interaction with the coupling strength $\epsilon_A=0.01$. The identical SL oscillators arrange themselves in two different clusters as per the vertex partitioning of the repulsive spanning tree. Additionally, these two groups possess a $\pi$ phase difference. Hence, we find $F_{chain}=0$. Despite the competitive interaction among those oscillators, the system still maintains its oscillatory states. The initial conditions are randomly drawn from $[-1,1] \times [-1,1]$. The value of $\epsilon_R=-0.1$ is selected from the saturated domain of $F$ for fixed $\epsilon_A=0.01$. The saturated numerically simulated value of $F$ is $0.2859$. Our theoretical prediction suggests $3-5$ attractive link will only contribute to the calculation of $F$, and thus our analytically predicted value of $F$ is $\dfrac{2}{7}$. Therefore, once again, our predicted value of $F$ matches with the numerically obtained value in the saturated domain.

\subsection{Another network with $\mathbf{6}$ nodes without containing the chain as a subgraph} \label{6_node_2}

\par For further validation, we consider another non-bipartite network with $N=6$ and $L=7$. This network does not possess the chain as its subgraph. In fact, although this network contains $8$ different spanning trees, still, none of these spanning trees can be decomposed into two disjoint sets of vertices $\{1,3,5\}$ and $\{2,4,6\}$. We select a spanning tree $1-2-4-6-5-3$ as highlighted by red arcs in Fig.\ \eqref{Picture15} (a). The vertex partitioning produces two independent and disjoint sets $V_1=\{1,4,5\}$ and $V_2=\{2,3,6\}$. We pass the repulsive coupling strength $\epsilon_R=-0.1$ through this spanning tree and pass the attractive coupling strength $\epsilon_A=0.01$  through the remaining chords (black arcs in Fig.\ \eqref{Picture15} (a)). For the theoretical prediction, we allocate the phase $\alpha$ to each oscillator of $V_1$ and the phase $\alpha+\pi$ to each element of $V_2$. The attractive link $4-5$ solely will contribute to the calculation of the measure $F$, and the predicted value of $F$ will be $\dfrac{2}{7}$. This predicted value fits well with our numerically simulated value $0.2859$ of $F$. Similarly, for the calculation of $F_{chain}$, the pair of oscillators ($2$,$3$) and ($4$,$5$) contribute. Thus, the predicted value of $F_{chain}=\dfrac{4}{5}$. Note that, although the graph in Fig.\ \eqref{Picture14} (a) does not contain the chain as its subgraph, still able to produce $F_{chain}=0$ for a suitable choice of the repulsive spanning tree. However, the network in Fig.\ \eqref{Picture15} (a) is never able to give zero value of $F_{chain}$ for our chosen spanning tree as well as for any other spanning trees too. The edges within the same sets $\{1,3,5\}$ and $\{2,4,6\}$ actually do not allow $F_{chain}$ to diminish to zero. However, the ensemble splits into two groups, and the members of each group are identified easily using the vertex partitioning $V_1=\{1,4,5\}$ and $V_2=\{2,3,6\}$ of the selected repulsive spanning tree. We plot the phases of each oscillator in Fig.\ \eqref{Picture15} (b) at a particular time after the initial transient. The position of the intrinsic phases confirms our successful prediction approach utilizing the vertex decomposition of the repulsive spanning tree. Figure \eqref{Picture15} (c) reflects the oscillatory nature of each element. Both these clusters are $\pi$ distance apart from one another as per Fig.\ \eqref{Picture15}.

%\section*{Appendix B}

\textcolor{black}{ 
\subsection{All spanning tress of the complete graph $\mathbf{K_4}$}\label{k4_spanning}
 There are $16$ spanning trees in total for the complete graph $K_4$. Below, We will list them by showing the edges present in each spanning tree. Each vertex is labeled as '1', '2', '3', and '4'.\\\\ Spanning trees possessing the sequence $(1,2,2,1)$ as the degrees of their vertices:
  \begin{enumerate}
      \item ( 1-2, 2-3, 3-4 ) (See subfigure (c) of Fig.\ \eqref{Picture3})
      \item ( 1-2, 1-3, 2-4 )
      \item ( 1-2, 1-4, 2-3)
      \item ( 1-3, 2-3, 1-4 )
      \item ( 1-4, 2-4, 2-3 ) 
      \item ( 1-3, 1-4, 2-4 )
      \item ( 1-2, 2-4, 3-4 )
      \item ( 1-3, 2-4, 3-4 )
      \item ( 1-2, 1-3, 3-4 )
      \item ( 1-2, 1-4, 3-4 )
      \item ( 1-3, 2-3, 2-4 )
      \item ( 1-4, 2-3, 3-4 )
  \end{enumerate}
Spanning trees possessing the sequence $(1,3,1,1)$ as the degrees of their vertices:
\begin{enumerate}
      \item ( 1-2, 1-3, 1-4 ) (See subfigure (b) of Fig.\ \eqref{Picture3})
      \item ( 1-2, 2-3, 2-4 )
      \item ( 1-3, 2-3, 3-4 )
      \item ( 1-4, 2-4, 3-4 )
\end{enumerate}
Each of these sets represents a spanning tree of the complete graph $K_4$. }

{ 
\subsection{Examining the Microscopic Dynamics of Fig.\eqref{Picture6}} \label{Fig6_micro}
 Upon close examination of the purple plots in Fig.\ \eqref{Picture6}, one can notice the distinctive spiky profile exhibited by the frustration order parameter $F$ in the neighborhood of $\epsilon_R=0$ due to multistabilty. This spiky profile raises the intriguing possibility of the presence of a hysteresis loop. Thus, to delve into the hysteresis loops, we focus on the complete network $K_4$ and select the repulsive spanning tree with edges $\{1-2, 1-3, 1-4\}$, akin to the purple plots depicted in Fig.\ \eqref{Picture6} of the manuscript. Our approach involves varying the coupling strength $\epsilon_R$ with a fixed step length of $0.00001$. During the forward transition, we vary $\epsilon_R$ from $0.01$ to $-0.03$, while for the backward transition, the range shifts from $-0.03$ to $0.01$. At the outset of both transitions, we set all initial conditions randomly from $[-1,1] \times [-1,1]$. Following the first iteration, we utilize the last simulation point as the initial conditions for subsequent iterations, with a minor perturbation added to each oscillator drawn from the interval $[0,0.01]$. In the absence of these perturbations, when the oscillators become synchronized, they maintain the same initial conditions in the subsequent iteration. Consequently, this synchronization inhibits the observation of any transition in $F$ during the forward transition. Thus, $F$ consistently remains at $2$ if we do not introduce these small perturbations. Hence, we opt not to change the value of $\epsilon_R$ adiabatically. Instead, we introduce small perturbations to each oscillator to disrupt synchronization and facilitate the observation of transitions.\\
Clearly, the coupling strength $\epsilon_R$ marking the forward transition ($F=2$ to $F<2$) differs from that of the backward transition, where $F=1$ to $F>1$ occurs. This distinction is evident in the subfigure (b) of Fig.\ \eqref{Picture105}, where we track the variation of the Kuramoto order parameter, denoted as $Order$. It is given by
$Order = \left| \frac{1}{N} \sum_{j=1}^{N} e^{i\theta_j} \right|$, such that $Order=1$ when the phases of the oscillators are completely synchronized. Despite multiple repetitions, the observed transition pattern remains consistent.\\
We have attempted to determine the forward and backward transition points analytically but currently find it challenging to do so. Furthermore, we have only analyzed this phenomenon within the context of Fig.~\ref {Picture6} and have yet to confirm its presence in other networks. Consequently, we hesitate to generalize or assert its robustness without further analytical tractability. It's worth noting that the parameter $F$ solely encompasses information about the phases of each oscillator, whereas we focus on amplitude oscillators in our study. Hence, conducting analytical simulations at this juncture poses challenges. However, we think
exploring this phenomenon more comprehensively can be an interesting future research endeavor.   
}
%\subsection*{COMPETING INTERESTS} The authors have no conflicts to disclose.
%
%
%\section*{AUTHORS’ CONTRIBUTIONS}
%
%
%{\bf Sayantan Nag Chowdhury:}  Conceptualization, Data curation, Formal analysis, Investigation, Methodology, Project administration, Resources, Software, Supervision, Validation, Visualization, Writing – original draft. %{\bf Francesco Sorrentino:} Formal analysis, Project administration, Supervision, Validation, Visualization, Writing - review \& editing. 
%{\bf Md.\ Sayeed Anwar:} Data curation, Formal analysis, Investigation, Project administration, Resources, Software, Validation, Visualization, Writing - review \& editing. {\bf Dibakar Ghosh:}  Data curation, Formal analysis, Project administration, Supervision, Validation, Visualization, Writing - review \& editing.

%\section*{DATA AVAILABILITY}
%Data sharing is not applicable to this article as no new data were created or analyzed in this study.

%	\section*{Acknowledgment}
%	S.N.C. would  like  to  acknowledge  the  CSIR  (Project No. 09/093(0194)/2020-EMR-I) for financial assistance.	

%\section*{DATA AVAILABILITY}
%The data that support the findings of this study are available within the article.\\
%

\typeout{}
\bibliographystyle{apsrev4-1}
\bibliography{biblo}

%\onecolumngrid

\end{document}